\documentclass[12pt]{article}
\usepackage{amsmath}
\usepackage{epsfig}
\usepackage{amsfonts}
\usepackage{amssymb}
\usepackage[dvipsnames]{xcolor}

\def\beq{\begin{equation}}
\def\eeq{\end{equation}}
\def\bea{\begin{eqnarray}}
\def\eea{\end{eqnarray}}
\def\beas{\begin{eqnarray*}}
\def\eeas{\end{eqnarray*}}
\def\Tr{{\rm Tr}}

\def\XXint#1#2#3{{\setbox0=\hbox{$#1{#2#3}{\int}$}
     \vcenter{\hbox{$#2#3$}}\kern-.5\wd0}}

\begin{document}


\begin{titlepage}

\begin{centering}

\vspace*{3cm}

{\Large\bf Remarks on the Parameter Dependence of QCD$_{2A}$}

\vspace*{1.5cm}

{\bf Uwe Trittmann}
\vspace*{0.5cm}

{\sl Department of Physics\\
Otterbein University\\
Westerville, OH 43081, USA}

\vspace*{1cm}

\today

\vspace*{2cm}

{\large Abstract}

\vspace*{1cm}

\end{centering}

Recently it was shown that there are only a few operators that
contribute to the (very weak) $N_c$ dependence of the spectrum of
two-dimensional adjoint QCD.
We provide numerical evidence that none of these operators is dominant.
In other words, all operators
are necessary to guarantee the faithful representation of
the Hamiltonian, resulting in, for instance,
exact degeneracies of certain multi-particle state masses.

We provide a complete list of interactions between single- and multi-trace
states, which differ by trace-number block, as expected from the
Kutasov-Schwimmer universality.
The completed list allows for a targeted deformation of
the theory to render some masses $N_c$-invariant. Nonetheless, we argue that
a small amount of $N_c$ dependence of the theory is unavoidable.

We also investigate the large constituent mass regime $m^2/g^2_{\rm YM}N_c\gg 1$
of the theory at large $N_c$ in the eLCQ approach. We find, among other things,
that as the
constituent mass grows, eigenstates get purer in parton number, in line
with expectations.

\vspace{0.5cm}


\end{titlepage}
\newpage


\section{Introduction}

Two-dimensional adjoint quantum chromodynamics (QCD)
has been an interesting testing ground for
mechanisms in full four-dimensional QCD for a long time.
The two-dimensionality allows one to discard
renormalization problems, but renders a dynamical gluon
impossible, which has led to the introduction of a fermion
transforming under a gluon-like adjoint representation.  
Most often, QCD$_{2A}$ is understood as a $G=SU(N)$ Yang-Mills theory coupled
to a single ($N_f=1$) Majorana fermion in the adjoint representation of
$G$ \cite{DalleyKlebanov,BDK,Kutasov94,GHK}, but lately results
have been generalized to arbitrary gauge groups
$G$ \cite{Gomis} and $N_f>1$, notably coupling to an adjoint Dirac fermion
($N_f=2$) in \cite{Schoutens} and most recently \cite{Damia2024}.
Like fundamental
QCD$_2$, i.e.~the 't Hooft model, adjoint QCD$_2$ is often tackled in
light-cone quantization and at large $N_c$ \cite{BDK,GHK}.  The theory
has several interesting features, like a confining-screening transition
in the massless limit\footnote{Ref.~\cite{GrossKlebMatSmilga} and for
  more recent developments \cite{Jacobson,KlebFundamental21,Smilga21}.}, and
asymptotic supersymmetry at $m^2_{\rm SUSY}=\frac{g^2_{\rm YM}N_c}{2\pi}$
\cite{Kutasov94,Boorstein}.

In the last few years, renewed interest in two-dimensional
adjoint QCD and related theories has lead to a wealth of new insights.
For instance, in Ref.~\cite{KlebFundamental21} fundamental Dirac fermions
were added to the theory, resulting in a $osp(1|4)$ symmetry algebra when all
fermions are massless. This allows for an alternative understanding of
degeneracies and emergence of continuum thresholds in the spectrum, which
can also be understood using the Kac-Moody algebra of the $SU(N)$ currents
\cite{KutasovSchwimmer}. Making the fundamentals massive yet keeping the
adjoints massless, one finds persistent exact degeneracies, showing
that the spectrum of mesons becomes continuous above a certain bound-state
mass. This in turn is strong evidence for the vanishing string tension in ---
and therefore screening behavior of ---
massless adjoint QCD$_2$. That a gapped theory like QCD$_{2A}$ is screening
in the massless limit is somewhat unusual, and had
been questioned in the wake of the
discovery \cite{Jacobson} that the adjoint theory possesses not only $N_c$
discrete $\theta$ vacua \cite{Witten79}, but also an exponentially (in $N_c$)
growing number of superselection sectors.
The analytic arguments in \cite{KlebFundamental21} regarding the
degeneracies of the spectrum of adjoint QCD$_2$ already clarified some
finite $N_c$ results discovered in \cite{Anton}. This was corroborated in a
full-on finite-$N_c$ assault on the theory the next year \cite{KlebFiniteN},
which exhibited the extremely weak dependence of the theory on $N_c$
and showed that the supersymmetry of the massive theory
(at $\pm m_{\rm SUSY}$) survives at finite $N_c$. 
More light on the supersymmetry of adjoint QCD$_2$ was recently shed via the
construction of a gauge invariant supercurrent \cite{KlebSupercurrents}. 

Generic light-cone quantization is based on a single, trivial Fock vacuum,
and therefore is unable to access the other $\theta$-sectors
of the theory's Hilbert space. This has motivated the deployment of
other non-perturbative approaches, such as the Hamiltonian lattice
formulation \cite{HamLattice1and2,DempseyGlueck}, standard
Euclidean lattice calculations \cite{Bergner2024},  
and conformal truncation \cite{Katz}.
In another line of work \cite{KlebSmallCircle}, adjoint QCD$_2$ was
put on a small spatial circle subject to {\em periodic} boundary conditions,
and the effective action was derived. The states can then be split into the
$\mathbb{Z}_N$ center symmetry sectors. At the supersymmetric mass (squared)
$m^2_{\rm SUSY}$, supersymmetry is spontaneously broken in all but the trivial
sector. Within the small-circle approximation it can also be shown
that for massless fermions there are $2^{N-1}$ sectors of the Hilbert space, and
that there is a large number of symmetry operators
(not necessarily invertible), in line with \cite{Seifnashri}. 

The study of two-dimensional adjoint QCD has thus fragmented into several
quite distinct lines of active research focusing on its many interesting
features. In this short note we will similarly elaborate on
several, not necessarily strongly connected open questions. 
The paper is then organized as follows.
We review the exhaustively-symmetrized Light-Cone Quantization (eLCQ)
approach in the next section.
In the following section,
we take a look at the constituent mass dependence of the theory
at large $N_c$. This is a straightforward check within the framework of
eLCQ, i.e.~using asymptotic eigenfunctions, that the theory behaves as
anticipated. Namely, we expect the eigenstates to become purer in parton
number, because pair creation is expensive due to the mass of the
constituents or partons.

Next we complete our work in \cite{UT6}, where we listed all interactions
starting or ending with a single-trace state, in an effort to understand the
insensitivity of the theory to changes in $N_c$. The argument put forth in
Ref.~\cite{UT6} was
that due to the singularity structure of QCD$_{2A}$, there are only a few
and very constrained
interactions possible --- which would explain the $N_c$ insensitivity.
Here, this argument is shown to be too naive.
While indeed matrix elements of several interactions are identically zero,
a few do only vanish where they must to avoid singularities,
for instance between states with identical
actor momenta. We complete our
characterization of interactions by evaluating the non-trivial interactions
between states with several traces. This highlights the
Kutasov-Schwimmer universality \cite{KutasovSchwimmer}, which stipulates
that sectors other than the identity and the adjoint block cannot
produce single-particle states, and thus must function differently.

The completed interaction table also allows us to emulate finite $N_c$
Discretized Light-Cone Quantization (DLCQ)
calculations and selectively change interaction strengths. We can thus
deform the theory, and find that the mass eigenvalues change
in an interesting pattern. In particular, the scheme seems to allow for
paths of $N_c$ invariance:
as one interaction becomes stronger, and another weaker, the bound-state
masses of some states remain invariant. Due to the many constraints
(multi-particle masses have to remain degenerate, etc.) we don't see a way
to turn the entire spectrum $N_c$ invariant.
We conclude with a discussion of our results.


\section{The eLCQ Framework}

\subsection{From the Lagrangian to the Light-Cone Hamiltonian}

Two-dimensional adjoint QCD with one Majorana fermion
is based on the Lagrangian
\beq\label{Lagrangian}
{\cal L}=Tr[-\frac{1}{4g^2_{\rm YM}}F_{\mu\nu}F^{\mu\nu}+
i\bar{\Psi}\gamma_{\mu}D^{\mu}\Psi],
\eeq
where $\Psi=2^{-1/4}({\psi \atop \chi})$, 
with $\psi$ and $\chi$ 
being $N_c\times N_c$ matrices. The field strength is
$F_{\mu\nu}=\partial_{\mu}A_{\nu}-\partial_{\nu}A_{\mu}+i[A_{\mu},A_{\nu}]$,
and the covariant derivative is $D_{\mu}=\partial_{\mu}
+i[A_{\mu},\cdot]$. The gamma matrices $\gamma^0=\sigma_2$ and
$\gamma^1=i\sigma_1$ are used.
As is customary, we formulate the theory using
light-cone coordinates $x^\pm=(x^0\pm x^1)/\sqrt{2}$,
where $x^+$ plays the role of a time.
We fix the gauge by $A^+=(A^0+ A^1)/\sqrt{2}=0$ (``light-cone gauge'').
The theory is described extensively in the literature, see 
\cite{DalleyKlebanov, BDK, GHK, KlebFiniteN} and references therein.

The dynamics of  QCD$_{2A}$ are described by its
light-cone momentum operator $P^+$ and Hamiltonian $P^-$.
Following the canonical procedure
via the energy-momentum tensor $\Theta^{\pm\pm}$,
we express these operators in terms of the dynamic fields. While the
constraint fields, i.e.~the gluon and the left-moving fermion, are eliminated,
the right-moving adjoint fermions $\psi_{ij}$ are quantized by imposing
anticommutation relations at equal light-cone times
\beq\label{PhiCR}
\left\{\psi_{ij}(x^{-}), \psi_{kl}(y^{-})\right\} = \frac{1}{2}\,
\delta(x^{-}-y^{-})\bigl ( \delta_{il} \delta_{jk}-
{1\over N_c}\delta_{ij} \delta_{kl}\bigr).
\eeq
The operators are then
\beas
P^+&=&\int dx^-\Theta^{++}=\frac{i}{2}\int dx^- \Tr\{\psi \partial_-\psi\},\\
P^-&=&\int dx^-\Theta^{+-}=-\frac{1}{2}\int dx^- \Tr\{im^2\psi
\frac{1}{\partial_-}\psi+\frac{g^2_{\rm YM}}{2} J^+\frac{1}{\partial^2_-}J^+\},
\eeas
where the $J^+_{ij}=\psi_{ik}\psi_{kj}$ are the right-moving components
of the SU(N) current.
We expand the fields in terms of fermion operators
\beq\label{PhiExpansion}
\psi_{ij}(x^-) = {1\over 2\sqrt\pi} \int_{0}^{\infty} dk^{+}
\left(b_{ij}(k^{+}){\rm e}^{-ik^{+}x^{-}} +
b_{ji}^{\dagger}(k^{+}){\rm e}^{ik^{+}x^{-}}\right ),
\eeq
with anti-commutation relations following from Eq.~(\ref{PhiCR})
\beq\label{Commy}
\{b_{ij}(k^{+}), b_{lk}^{\dagger}(p^{+})\} =
\delta(k^{+} - {p}^{+})
(\delta_{il} \delta_{jk}-\frac{1}{N_c}\delta_{ij} \delta_{kl})\,\,.
\eeq
The dynamics operators of QCD$_{2A}$ are
\begin{eqnarray}\label{ModeDecomp}
P^+ &=& \int_{0}^{\infty} dk\ k\, b_{ij}^{\dagger}(k)b_{ij}(k)\ ,\\
P^{-} &=& {m^2\over 2}\, \int_{0}^{\infty}\nonumber
{dk\over k} b_{ij}^{\dagger}(k)
b_{ij}(k) +{g_{Y\!M}^2 N\over 2\pi} \int_{0}^{\infty} {dk\over k}\
C(k) b_{ij}^{\dagger}(k)b_{ij}(k) \\
&&+ {g_{Y\!M}^2\over 4\pi} \int_{0}^{\infty} dk_{1} dk_{2} dk_{3} dk_{4}
\biggl\{ B(k_i) \delta(k_{1} + k_{2} +k_{3} -k_{4})\nonumber \\
&&\qquad\qquad\times(b_{kj}^{\dagger}(k_{4})b_{kl}(k_{1})b_{li}(k_{2})
b_{ij}(k_{3})-
b_{kj}^{\dagger}(k_{1})b_{jl}^{\dagger}(k_{2})
b_{li}^{\dagger}(k_{3})b_{ki}(k_{4})) \nonumber\\
&&\qquad + A(k_i) \delta (k_{1}+k_{2}-k_{3}-k_{4})
b_{kj}^{\dagger}(k_{3})b_{ji}^{\dagger}(k_{4})b_{kl}(k_{1})b_{li}(k_{2})
 \nonumber\\
&&\qquad + \frac{1}{2} D(k_i) \delta (k_{1}+k_{2}-k_{3}-k_{4})
b_{ij}^{\dagger}(k_{3})b_{kl}^{\dagger}(k_{4})b_{il}(k_{1})b_{kj}(k_{2})
\biggl\}, \nonumber
\end{eqnarray}
with
\begin{eqnarray}\label{HamFunctions}
A(k_i)&=& {1\over (k_{4}-k_{2})^2 } -
{1\over (k_{1}+k_{2})^2}\ , \label {ATerm} \\
B(k_i)&=& {1\over (k_{2}+k_{3})^2 } - {1\over (k_{1}+k_{2})^2 }, \nonumber\\
C(k)&=& \int_{0}^{k} dp \,\,{k\over (p-k)^2},\nonumber \\ 
D(k_i)&=& \frac{1}{(k_{1}-k_{4})^2} - \frac{1}{(k_{2}-k_{4})^2}\nonumber.
\end{eqnarray}
It will be useful to name the interactions by their letter ($A,B,C,D$) followed
by an integer $n$ indicating their suppression at finite $N_c$, namely
$1/N_c^n$. Additionally, we
use a tilde (e.g.~$\widetilde{A1}$) to mark trace-splitting interactions, a
bar (e.g.~$\overline{B2}$) for trace-joining interactions, and underline
trace-diagonal interactions (e.g.~$\underline{A0}$). The structure of the
Hamiltonian is
\[
P^-=P^-_m+P^-_{\rm ren}+P^-_{\rm PC,s}+P^-_{\rm PC,r}+P^-_{\rm PV}+P^-_{{\rm finite} N},
\]
with mass, renormalization, parton-number conserving singular and regular
operators, as well as parton-number violating and finite $N_c$ operators. 

Note that the Hamiltonian is symmetric under reversal of indices
(``charge conjugation'')
\beq\label{Tsymmetry}
{\cal T}: b_{ij} \to b_{ji},
\eeq
so we can split the spectrum into its ${\cal T}$ even and odd components
with quantum numbers $T=\pm 1$.
This symmetry is very useful.
Originally introduced in \cite{Kutasov94}, it has been used in
\cite{KlebFiniteN} as an outer automorphism of the current algebra
to find single-particle bound states buried in a multi-particle continuum.

\subsection{The eLCQ Method}

We can now formulate the eigenvalue problem
\[
2P^+P^-|\Psi\rangle=:H_{LC}|\Psi\rangle=M^2|\Psi\rangle
\]
for the masses (squared) $M^2$ and the eigenstate $|\Psi\rangle$
of the light-cone Hamiltonian $H_{LC}$.
For now, let's consider the large $N_c$ theory, and review the approach
put forth in \cite{UT3and4,UT5}, namely
{\em exhaustively-symmetrized Light-Cone Quantization} (eLCQ).
We start with the observation that
the theory in the asymptotic regime, i.e.~for states with large bound-states
masses $M^2$,
naturally decomposes into sectors of definite parton number $r$,
and the full, large $N_c$ eigenstate can be written
\beq\label{TheStates}
|\Psi\rangle=\sum_{r=2}^\infty|\Psi_r\rangle,
\quad\mbox{ with }
|\Psi_r\rangle=
\int\frac{d[x_r]}{{N_c^{r/2}}}\psi_r(x_1,x_2,\ldots,x_r)
Tr[b(-x_1)\cdots b(-x_r)]|0\rangle.
\eeq
The rather complicated integral limits of
\beq\label{IntMeasure}
\int d[x_r]:=
\int_0^{\frac{1}{r}} dx_1
\left(
\prod^{r-1}_{j=2}
\int^{1-(r-j+1)x_1-\sum^{j-1}_{k=2}x_k}_{x_1}dx_j
\right)
\eeq
define the {\em unique Hilbert space}
(uHS). In comparison with Eq.~(4.3) of \cite{Kutasov94}, they
replace the momentum-conserving delta function.

To solve the eigenvalue problem at large $N_c$, we expand the unknown
wavefunctions $\psi_r$ in terms of the solutions $\phi_r$ of the {\em
  asymptotic} problem, defined by the Hamiltonian
\[
H_{asymp}:=P^-_{PC,s}+P^-_{ren},
\]
which has just the most singular part of the full Hamiltonian --- plus the
renormalization term to render it finite in the principal value sense.
For simplicity\footnote{In principle, it is
  possible to write down an analytic solution even without this approximation,
  see Ref.~\cite{Komatsu} for the 't Hooft model, and \cite{Asrat} for the
  two- and three-parton solutions of QCD$_{2A}$.}
 we make the additional assumption that the interaction
integral over the kernel $K(x-y)$ is dominated by the region where $x\approx y$.
For example,
\beq\label{approx}
\int^1_0\frac{\phi(y)dy}{(x-y)^2} \approx 
\int^\infty_{-\infty}\frac{\phi(y) dy}{(x-y)^2}.
\eeq
Then the simplest solution of the integral equation is a pure phase $e^{i\pi nx}$
with an integer {\em excitation number} $n$ leading to
an ansatz for the generic solution 
\beq\label{SimpleAnsatz}
\prod_{j=1}^{r-1}e^{i\pi n_j x_j}
=e^{i\pi \sum_{j=1}^{r-1}n_j x_j}=: e^{i\pi \vec{\nu}\cdot\vec{x}},
\eeq
where the constraint momentum ($\sum_i x_i=1$)
compels us to operate on a $(r-1)$-hyperplane
of the naive $r$-dimensional Hilbert space of $r$ parton states, which
is represented by the ``Greek vector'' $\vec{\nu}$.
An eigenfunction of the Hamiltonian will then be a sum of such
exponential terms, written for now symbolically as  
\[
\phi_{\vec{\nu}}(\vec{x})=\phi_{\vec{\nu}}(x_1,\ldots, x_r) =
\sum_{{\cal G}\vec{\nu}} (-1)^{s(\vec{\nu})} e^{i\pi
  \vec{\nu}\cdot\vec{x}},
\]
where the sign of the term is determined via $s(\vec{\nu})$ by group
theoretical arguments, and the sum is over all permutations $g\in{\cal
  G}$ allowed by the symmetries of the theory.

The eigenfunctions of the theory can then be expanded in the asymptotic basis
\[
\psi_r=\sum_{\vec{\nu}} a_{r,\vec{\nu}}\phi_{r,\vec{\nu}},
\]
where the sum is over all {\em bona fide} asymptotic states of the
theory, which can be written down in the approximation,
Eq.~(\ref{approx}), as a linear combination of sines and cosines. For
instance, the four lowest three-parton $T+$ states are \bea
\phi_{3+}^{(n,m)}(x_1,x_2,x_3)&=& \sum_{i=1}^{3} \cos (\pi n x_i + \pi
m x_{i+1}) + (n \leftrightarrow m), \eea with
$\vec{\nu}=(n,m)=(0,0),(2,2),(4,2),(4,0)$.
The excitation numbers are in general not unique, e.g. $(4,2)=(2,-2)$, so we
pick a representative.
Note that these are states in the $T_s=-T$ sector, and that in general
$T_s=(-1)^{r/2}T$ for even and $T_s=(-1)^{(r-1)/2}T$ for odd parton numbers $r$,
so that $T_s=T$ for $4n$ and $4n+1$ partons, $n\in\mathbb{N}$.

In practice, we want to
calculate Hamiltonian matrix elements
\beq\label{MatrixElements}
\langle \phi_{\vec{\mu}}|\hat{H}|\phi_{\vec{\nu}}\rangle
=\int_{uHS}\!\!d\vec{x}\int_{uHS}\!\!d\vec{y}\,\,\,
\langle \phi_{\vec{\mu}}|\vec{x}\rangle\langle\vec{x}|\hat{H}
|\vec{y}\,\rangle\langle\vec{y}\,|\phi_{\vec{\nu}}\rangle,
\eeq
and solve the matrix equation\footnote{It is a finite matrix equation
  when the number of basis states is cut off, $N_{\phi}<\infty$.}
\beq\label{ABFEVP}
\langle \phi_{\vec{\mu}}|\hat{H}|\phi_{\vec{\nu}}\rangle
=M^2\langle \phi_{\vec{\mu}}|\phi_{\vec{\nu}}\rangle.
\eeq
for the mass (squared) eigenvalue $M^2$.
Since the states have definite parity under cyclic permutations ${\cal C}$,
(as well as inversions ${\cal I}$,  and lower-dimensional inversions ${\cal S}$
 \cite{UT5})
we can simplify the evaluation of the matrix elements as seen in the
following example. Consider the matrix elements of the
most singular part of the Hamiltonian
relative to a single-momentum basis \cite{UT5} ($\vec{x}_{i,Sp}$ and $\vec{y}_{j,Sp}$
are spectator momenta)
\beas
\langle \vec{x}|\hat{H}_{PC}|\vec{y}\,\rangle&\supset&-\frac{g^2_{\rm YM}N_c}{2\pi}
\sum_{i=1}^{r'}\sum_{j=1}^r\frac{(-1)^{(r+1)(i+j
  )}}{(x_{i}-y_{j})^2}
\delta(x_i+x_{i+1}-y_j-y_{j+1})\delta(\vec{x}_{i,Sp}-\vec{y}_{j,Sp}).\\
\eeas
Notice that the sums explicitly symmetrize the expression under ${\cal C}$.
From this observation it is intuitive that the matrix element can be
evaluated alternatively by dropping the sums and instead enlarging the
integration domain to
\[
\mbox{cHS}:=\mbox{uHS}\cup{\cal C}\mbox{(uHS)}\cup{\cal C}^2\mbox{(uHS)}
+\ldots\cup{\cal C}^{r-1}\mbox{(uHS)}.
\]
To wit
\beq\label{IntcHS}
\int_{uHS}\!\!d\vec{x}\int_{uHS}\!\!d\vec{y}\,\,\,
\langle \phi_{\vec{\mu}}|\vec{x}\rangle\langle\vec{x}|\hat{H}
|\vec{y}\,\rangle\langle\vec{y}\,|\phi_{\vec{\nu}}\rangle=
\int_{cHS}\!\!d\vec{x}\int_{cHS}\!\!d\vec{y}\,\,\,
\langle \phi_{\vec{\mu}}|\vec{x}\rangle\langle\vec{x}|\hat{H}
|\vec{y}\,\rangle\big|_{i=1\atop j=1}\langle\vec{y}\,|\phi_{\vec{\nu}}\rangle.
\eeq
Incidentally, this simplifies the integral limits, and we can easily calculate
the integration volume 
\beq\label{CellVolume}
\int_{cHS} \!\!d\vec{x}=\int_0^1dx_1\prod_{j=2}^{r-1}\int_0^{1-\sum^{j-1}_{k=1}x_k}dx_j
=\frac{1}{(r-1)!},
\eeq
which makes sense since the {\em exhaustive set} ${\cal E}$ of symmetries
\cite{UT5} plus identity has $\frac{1}{2}(r-1)!$ members, with the order of the
full
symmetry group $|{\cal G}|=|{\cal C}\otimes{\cal I}\otimes{\cal T}\otimes{\cal E}|=2r!$ \cite{UT5}.


We now generalize to finite $N_c$.
Consider states at large $N_c$, but with more than one trace
\beq\label{MTrFockStates}
|\Psi^{(t)}_{\vec{r}}\rangle\!=\! \frac{1}{N_c^{r/2}}\!\int_0^1 \!\!dx_1
\prod_{j=2}^{r-1} \int_0^{1-\sum_{k=1}^{j-1} x_k} \!\!\!\!\!\! dx_j
\prod_{s=1}^t\psi_{r_s}\!\!\left(x^{(s)}_{1},x^{(s)}_{2},\ldots, x^{(s)}_{r_{s}}\right)
\!\Tr\!\left\{b^{\dagger}(x^{(s)}_{1})\cdots b^{\dagger}(x^{(s)}_{r_{s}})\right\}\!|0\rangle,
\eeq
where $x^{(s)}_{r_s}$ is the $r_s$th and
therefore ``last'' momentum in the $s$th trace.
As an example,
\beas
|\Psi^{(2)}_{2,3}\rangle&=&\frac{1}{N_c^{5/2}} \int_0^1 dx_1\int_0^{1-x_1}
\!\!\! dx_2\int_0^{1-x_1-x_2} \!\! \!\!\!dx_3
  \int_0^{1-x_1-x_2-x_3} \!\! \!\!\! \!\!\!dx_4\,\,
\psi_{2}(x_1,x_2)\psi_{3}(x_3,x_4,x_5) \\
&&\quad\qquad\qquad\qquad\qquad\qquad\qquad
\times\Tr\{b^{\dagger}(x_{1})b^{\dagger}(x_{2})\}
\Tr\{b^{\dagger}(x_{3})b^{\dagger}(x_{4})b^{\dagger}(x_{5})\}|0\rangle
\eeas
is a fermionic two-trace, multi-particle, five-parton
state made from a two-parton boson and a three-parton fermion.

These are eigenstates of the Hamiltonian only at large $N_c$. For finite $N_c$
they will couple to other single- or multi-trace states. The eigenvalue problem
is
\[
H_{LC, finite N}|\Phi\rangle=M^2|\Phi\rangle,
\]
where the eigenfunctions are linear combinations of all large $N_c$ of arbitrary
trace number
\[
|\Phi\rangle=\sum_{t=1}^\infty\sum_{\vec{r}} c_{t,\vec{r}}|\Psi^{(t)}_{\vec{r}}\rangle,
\]
because of the mixing due to trace-number violating operators $A1,B1,B2_{3Tr},D1$,
with a trace-splitting and a trace-joining version each \cite{UT6}. We will see below that there
are also two new, truly multi-trace interactions, $\widetilde{B3}_{23}$ and  $\overline{B3}_{32}$.
The operator notation defined below Eq.~(\ref{HamFunctions})
originated in Ref.~\cite{UT6} for one and two-trace
states, where it was also shown that these are the only interactions with
non-zero matrix elements relative to a single-momentum basis.


\section{Eigenfunctions at large constituent mass}

We saw that the eLCQ method is based on the assumption of dominance of
the singular term in the interaction, and that these asymptotic
solutions are linear combinations of trigonometric functions with
$r-1$ excitation number (which can be interpreted as positions\cite{Yung}).
The opposite limit is $m^2\rightarrow \infty$, which is
equivalent to the free theory $g^2_{\rm YM}N_c\rightarrow 0$. In this limit we
have a diagonal Hamiltonian, as only the mass term and the
renormalization term survives, so we should find delta-peaks as
eigenfunctions in momentum space (definite single-particle
momenta $x_i$). So far, attempts to find eigenfunctions in a regime
where $m^2$ is large compared the the coupling but still much smaller
than the bound state mass
\[
g^2_{\rm YM}N_c\ll m^2 \ll M^2
\]
have proven elusive \cite{Aharony}. 
Typically, one finds that most bound states with $r$ partons
have masses comparable to the bound state mass \cite{Asrat}
\[
r m^2\approx M^2.
\]
In this regime corrections are large, and one cannot get a good approximation
of the bound state masses.
On the other hand, for very heavy quarks, essentially assuming that there
is no coupling between sectors of different parton number, one can derive
exact eigenstates, at least for the first few parton sectors via an $r$-body
Schr\"odinger equation \cite{Asrat}. For four or more partons,
the ${\cal S}$-symmetry constraint \cite{UT5} is no longer trivial, and one
has a harder time deriving an appropriate Schr\"odinger equation.

%
\begin{figure}[ht]
\centerline{
\psfig{file=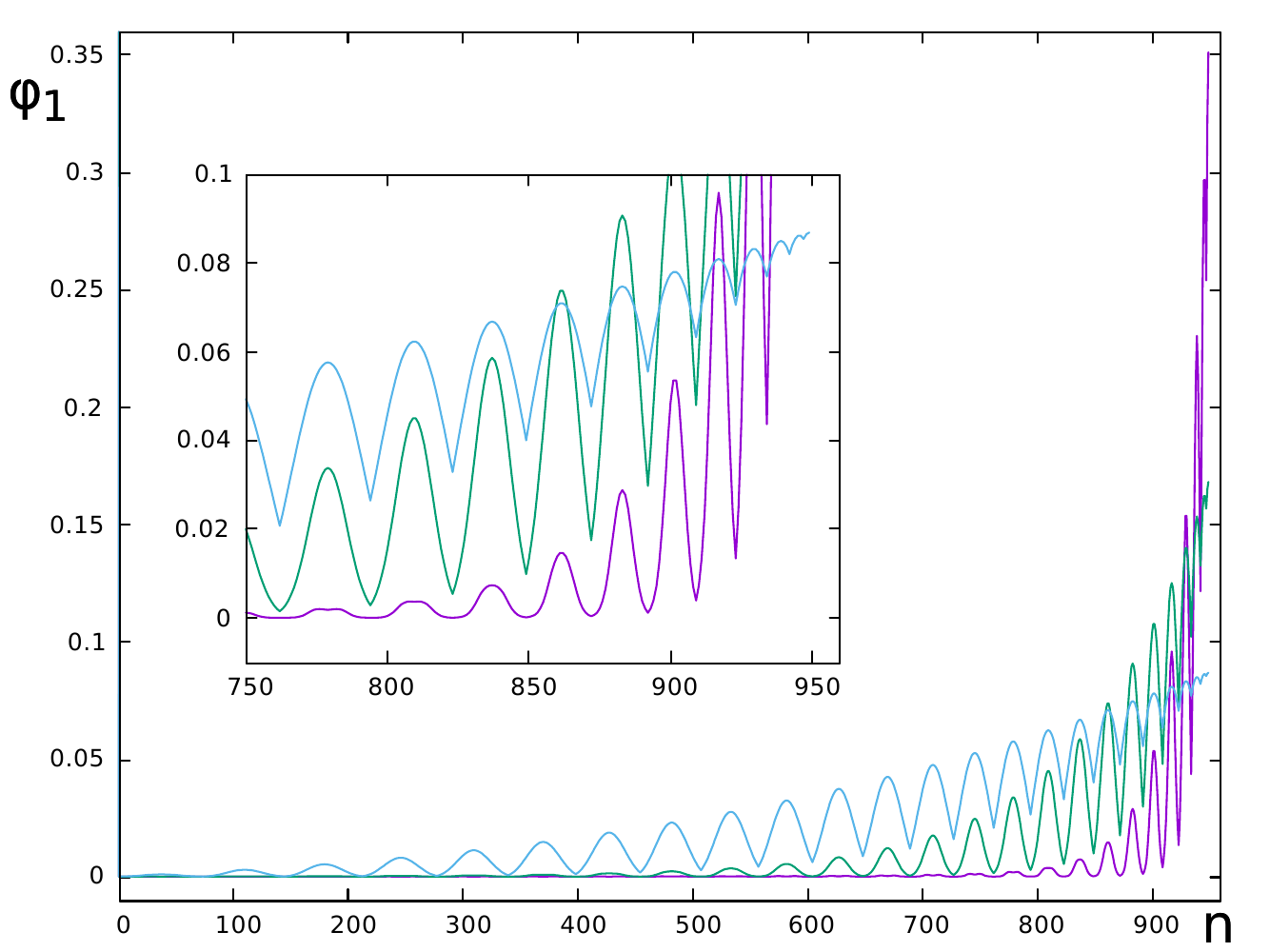,width=8cm}
\psfig{file=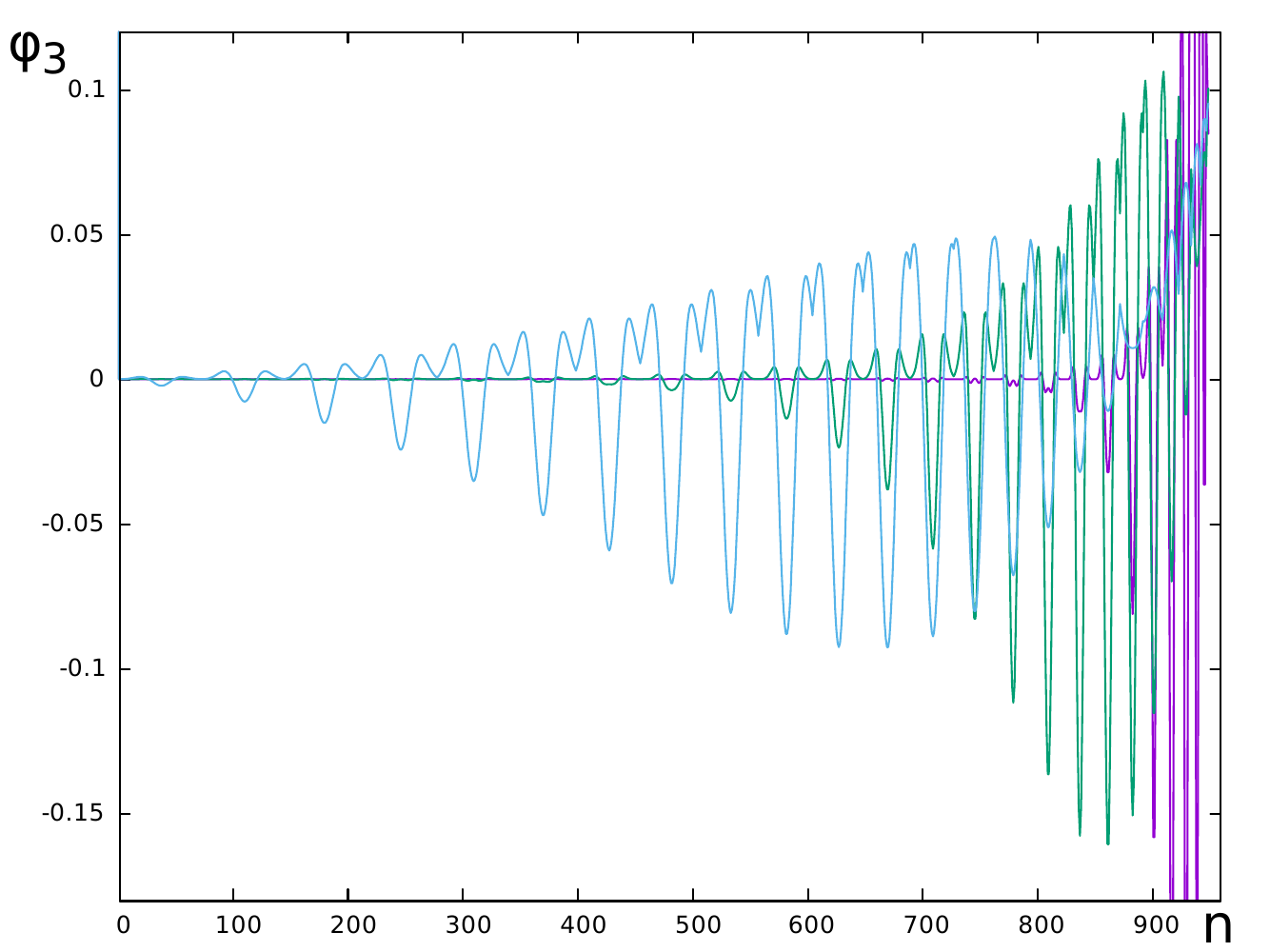,width=8cm}
}
\caption{Plots of the first (left) and third (right) eigenfunction at different
  constituent masses $\mu=10, 100, 1000$ (with darker shades of colors)
  as a function of the basis state number $n$.
  It shows a tendency to form a peak at
  large basis state number, which are states with about equal single-particle
  momenta, while
  still conforming to system constraints. The inset zooms in on the
  high-basis-number regime.
}
\label{FigsEFsMass}
\end{figure}
%

Looking at the numerical eigenfunctions, Fig.~\ref{FigsEFsMass}, we see
that as the constituent mass grows, states
with similar individual momenta carry more weight in the eigenfunctions.
The eigenfunctions at any $m^2<\infty$ are still subject to the symmetry
constraints,
so their maxima and minima are at fixed positions (e.g.~where the
wavefunctions have to vanish). With only the mass term we do not
expect the constraints to be enforced, and a featureless delta-peak emerges,
as we checked.

%
\begin{figure}[ht]
\centerline{
\psfig{file=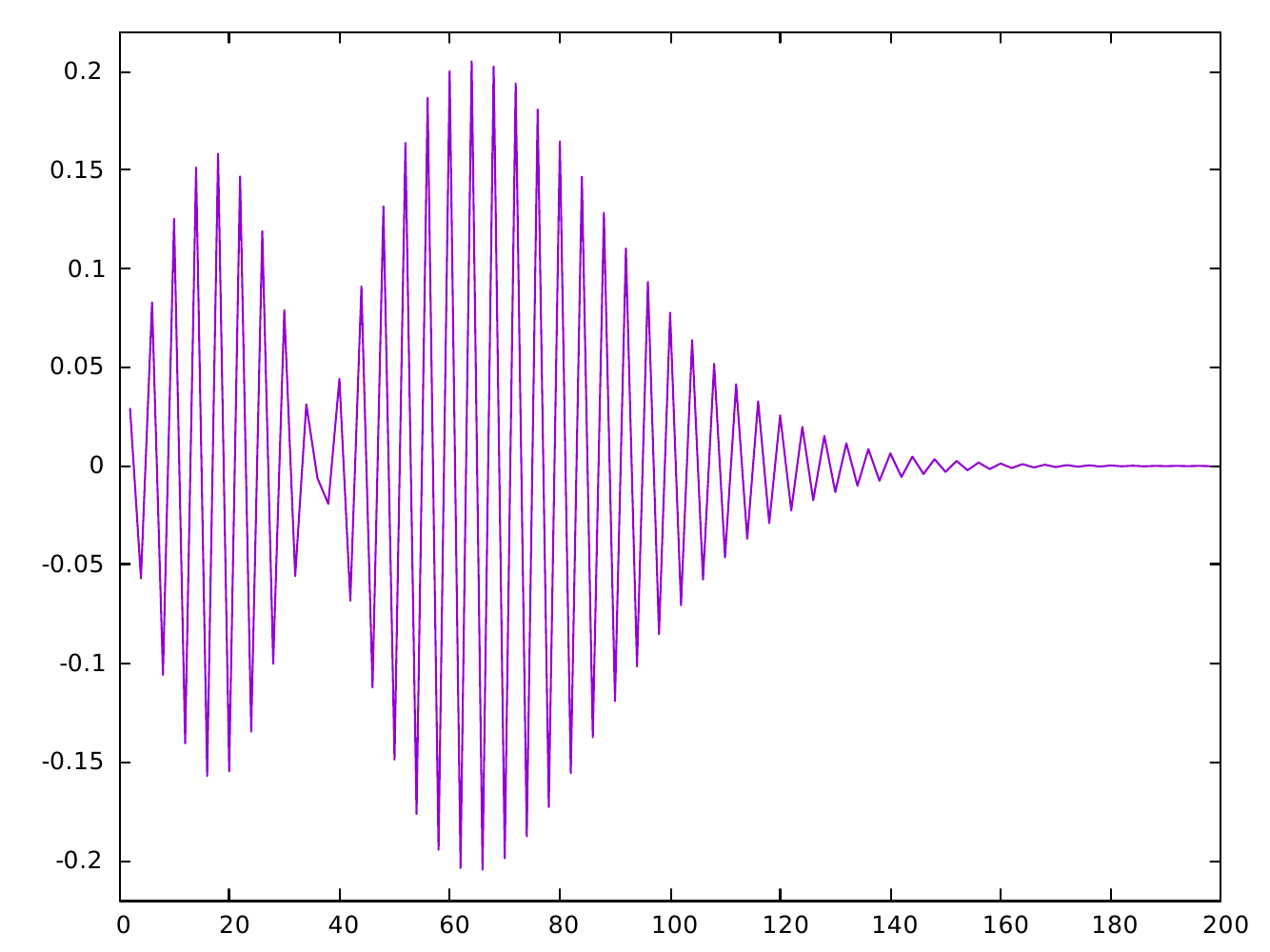,width=10cm}
}
\caption{Plot of the expansion of the second
  two-parton eigenstate at mass $\mu=10^5$ in terms
of two-parton asymptotic eigenstates $\varphi_{2,i}, i=1,2,\dots 200$.}
\label{FigsEFsMass2}
\end{figure}
%

Intuitively, the shape of a two-parton eigensolution 
is that of a Fourier transform of a delta peak. In other words, the delta peak,
i.e.~the infinite constituent mass eigenfunction, is expressed as a linear
combination of sines of shorter and shorter wavelength.
However, due to the symmetries or rather constraints
$({\cal C},{\cal I},{\cal S}),$ in the system, the only allowed (Fourier) modes
are the {\em bona fide} asymptotic eigenstates $\phi_{\vec{\mu}}$.
Since $n\leftrightarrow -n$ is always a symmetry, 
the generic decomposition of eigenfunctions looks like 
Fig.~\ref{FigsEFsMass2}.
At $m^2=\infty$ we obtain a symmetric curve $\pm\sin(2\pi n/n_B)$, where $n_B$
is the number of basis states, and $n\in[1,n_B]$. In other words,
the decomposition of a delta peak is a single-wavelength sine as expected.

Let's now look at the parton-number
purity of states for growing constituent mass.
This is a straightforward application of the eLCQ framework \cite{UT3and4}.
Once the asymptotic basis functions have been determined, and the
matrix elements have been computed and stored independently for the salient
parts of the Hamiltonian (singular, regular, parton-number violating, and
mass term), we can set up a finite-dimensional matrix
for any values of parameters, most notably the mass parameter
$\mu:=m^2\frac{2\pi}{g^2_{\rm YM}N_c}$, at
negligible numerical cost and solve for the eigensystem.

Parton purity in eLCQ is basically synonymous with convergence,
since the finite parameters in the eLCQ approximation are the
maximal parton number and the number of basis states in each parton
sector.  All should be sent to infinity in the continuum
limit. To run up parton number is numerically exceedingly costly \cite{UT5},
so we
are limited to $r<9$, whereas the number of basis states typically does
not have a big influence once it is reasonably large.

\begin{figure}[ht]
\centerline{
\psfig{file=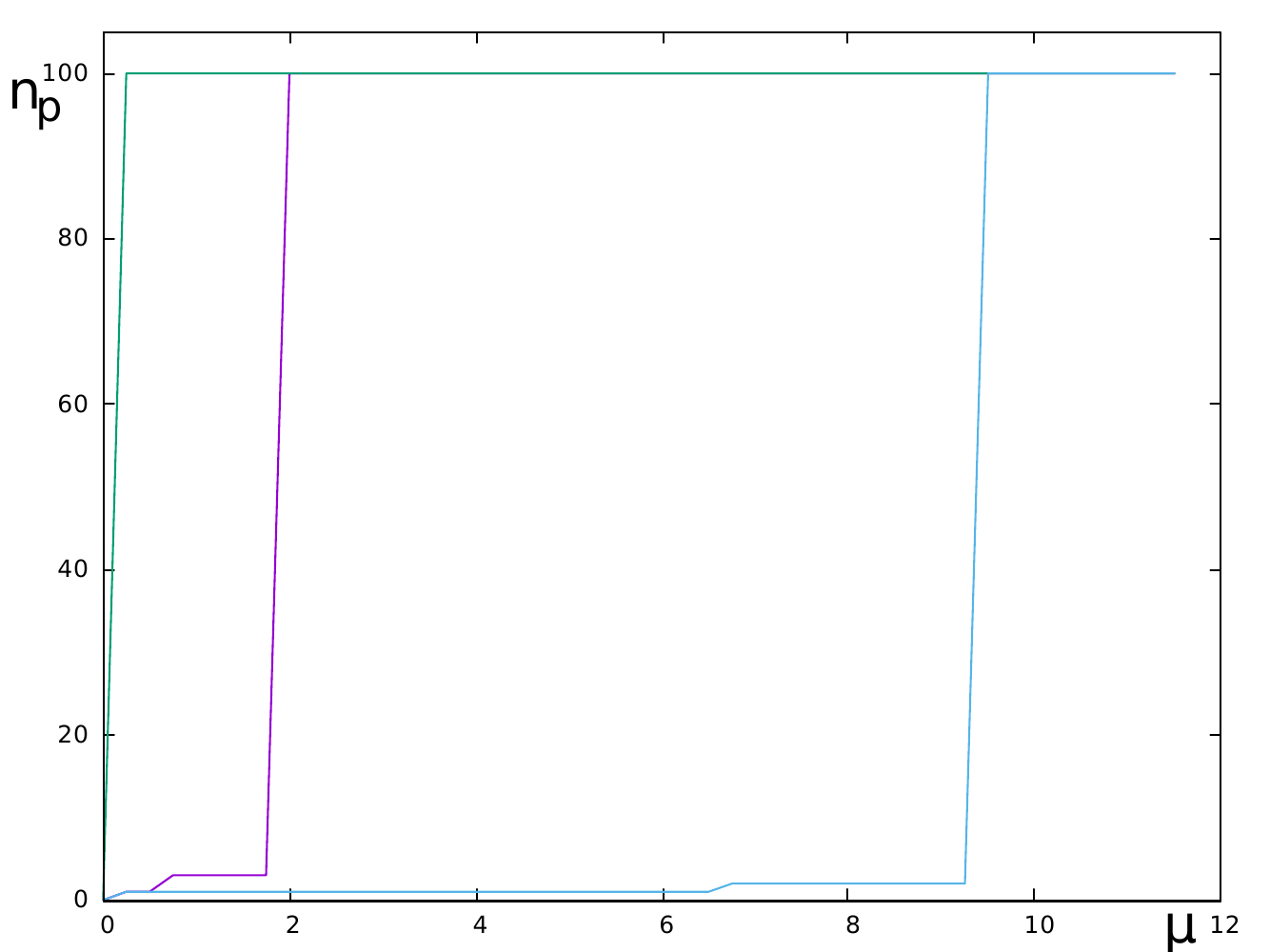,width=8cm}
\psfig{file=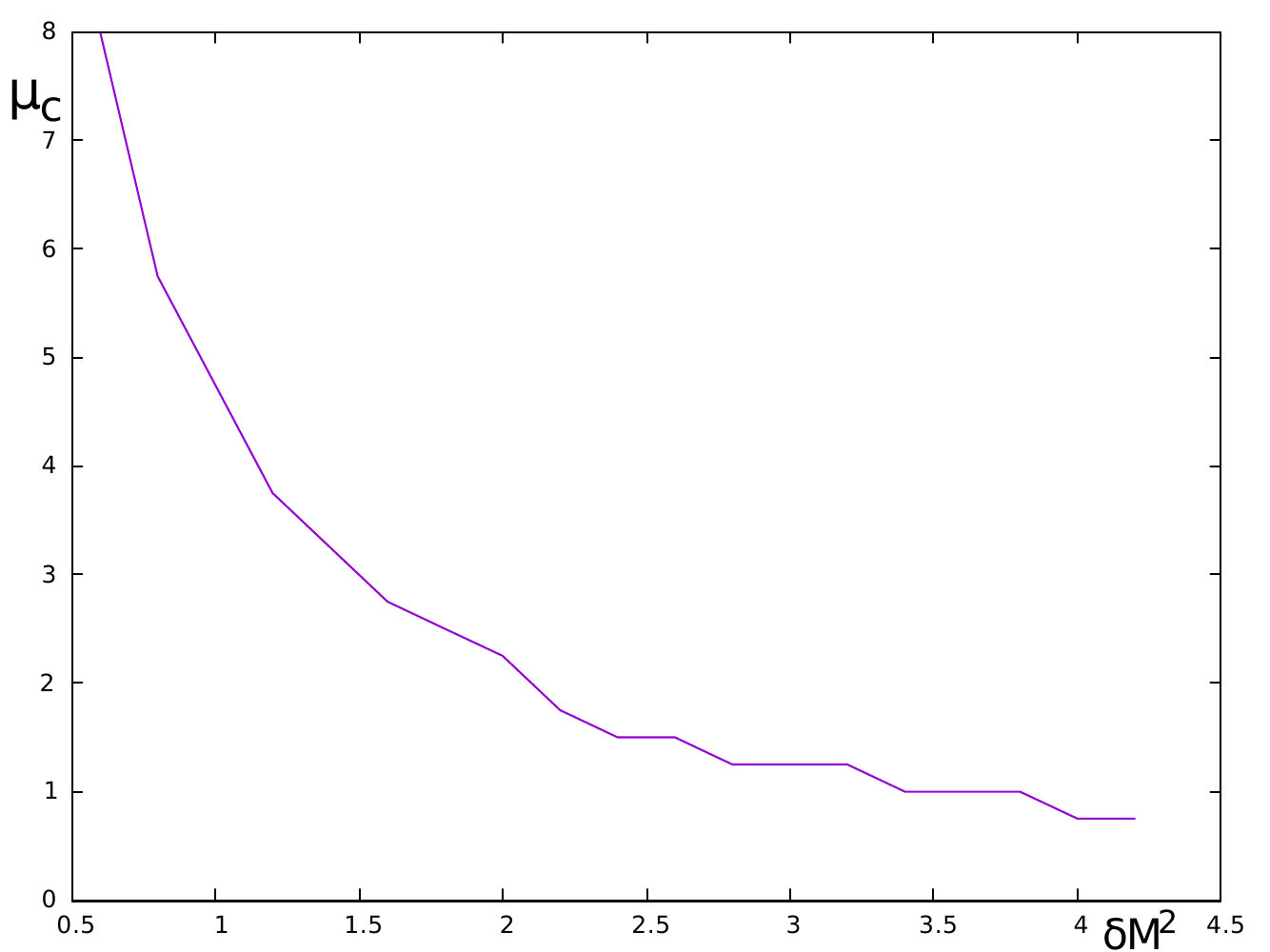,width=8cm}
}
\caption{(a)
  Number of parton-pure eigenstates $n_p$ as a function of $\mu$ in the
  fermionic $T_s$ odd sector (left). 
  The curves were computed with different
  criteria of purity, namely $\delta M^2 < 0.2\%, 1\%, 5\%$,
  with increasing critical mass parameter $\mu_{c}$
  above which all states are pure.
  (b) Plot of $\mu_{c}$ as a function of the criterion parameter
  $\delta M^2$ in the bosonic $T_s$ even sector (right). 
  In both cases 100 states were used in the lowest parton sector.}
\label{NrPureStates}
\end{figure}

One measure for convergence could be the percent difference between consecutive
parton number runs, for instance
between a run with the minimal number and one with
maximal parton number. Let's call this $\delta M^2$.
Then pick some threshold, say $\delta M^2<1\%$, which allows us to pick
a mass (squared) threshold up to which states are converged by this definition.
Overall we find that the number of pure states grows abruptly with the
constituent mass $m^2$, see Fig.~\ref{NrPureStates}(a). It is obvious
that the number of pure states shoots up to the maximal possible number at a
``critical'' mass parameter $\mu_{c}$, which depends on the criterion used,
see Fig.~\ref{NrPureStates}(b).
This qualitative behavior is independent of the exact criterion used, the
same in all sectors, and is consistent with the
expectation that the number of pure states grows exponentially with the
constituent mass \cite{KlebPrivate}.
These numbers should be taken with a grain of salt;
the counting of pure states is
delicate. If we use the plausible rule to ignore all states with masses
larger than the first mass that is off by more than the
criterion\footnote{High-mass states often receive corrections only from states
  with rather high parton number $r$, and would therefore be
  erroneously identified as {\em pure} with a low $r$ cutoff in place.}
$\delta M^2$, it is often the ``fault''  
of {\em one} impure state that
the higher mass states are discarded, even though 
their {\em percent} discrepancy, lowered by their higher mass, is
below the criterion $\delta M^2$.
Nonetheless, it seems safe to say that qualitatively, we have an
abrupt increase of pure states as the constituent mass increases, and that
many states are pure when this mass is still moderately small.

%
%


\section{True Multi-Trace Interactions}

We now switch gears and consider the finite $N_c$ theory.
The interactions of QCD$_{2A}$ listed in Ref.~\cite{UT6} are between
single- and multi-trace
states. At finite $N_c$ we find also interactions between
two multi-trace states. Some of them will be trivial, in the
sense that only one of the traces of one state interacts with one
or several traces of the other state, while there are spectator traces on both
sides.
It turns out that there are other, {\em true} multi-trace interactions, which
are therefore not part of the list in Ref.~\cite{UT6}.
Since we have at most three annihilators which can engage three traces
simultaneously, the most complicated (true) interaction is 
$3\Tr\leftrightarrow 2\Tr$, which is suppressed
by $1/N_c^3$, so name it $\widetilde{B3}_{23}$ or $\overline{B3}_{32}$, depending
on whether it splits or joins traces.
We find three additional true multi-trace interactions:
$\underline{A2}_{22},$
$\underline{D2}_{22},$
$\underline{B2}_{22}$.
The last three are trace diagonal.

Let's look at these five interactions in more detail.
Note that we can always cyclically rotate to act on the first parton in a trace.
The $\underline{A2}_{22}$ term comes from 
a sequence of terms generated by two annihilations,
namely
\[
\left(-\frac{1}{N_c}\delta^{a_1}_{a_2}\delta^{i}_{l}\delta_{k_2}^{y^{(1)}_1}\right)
\left(\delta^{b_1}_{k}\delta^{b_2}_{l}\delta_{k_1}^{y^{(2)}_1}\right),
\]
where the first, non-planar factor is due to the annihilator $b_{il}(k_2)$
acting on the first parton of the first trace with momentum $y^{(1)}_1$.
This is a $1/N_c^2$ contribution, with $1/N_c$ from the non-planar factor, and
$1/N_c$ from the missing $N_c$ in the term's coupling.
One parton from each trace is
replaced by two partons in just one of the traces. This is a trace-diagonal
interaction
\beq\label{A222D222}
\underline{A2}_{22}: |y_u\cdots y_{u-1}\rangle |z_w, \cdots z_{w-1}\rangle\to
|y_{u+1}\cdots y_{u-1}\rangle|k_3,k_4, z_{w+1}\cdots z_{w-1}\rangle.   
\eeq
The $D$-term version replaces the parton from each trace by two
adjacent partons in one trace with term sequence
\[
\left(-\frac{1}{N_c}\delta^{a_1}_{a_2}\delta^{k}_{j}\delta_{k_2}^{y^{(1)}_1}\right)
\left(\delta^{b_1}_{i}\delta^{b_2}_{l}\delta_{k_1}^{y^{(2)}_1}\right),
\]
which leads to the same expression, Eq.~(\ref{A222D222}), and interaction
diagram (see
below) as $\underline{A2}_{22}$, but carries a different function of the
momenta.
The adjacency is forced by the non-planar term which couples the
``inner'' indices of the new partons, $j=k$.
Since they have the same structure, a cancellation of
these divergent terms is possible. The regular $A$ term survives. 

For $\underline{B2}_{22}$ we find the following five {\em different} patterns.
\begin{itemize}
\item Partially adjacent: Annihilators adjacent, partons
  non-adjacent plus parton
  of other trace can be either $k_1$ or $k_3$ annihilator (PA1 \& PA3)
  \[
  |y_u\cdots y_{u-1}\rangle |z_w\cdots z_{s}\cdots z_{w-1}\rangle\to
  |z_{w+1}\cdots z_{s-1}\rangle |k_4,z_{s+1}\cdots z_{w-1},y_{u+1}
  \cdots y_{u-1}\rangle,
  \]
  \[
  |y_u\cdots y_{u-1}\rangle |z_w\cdots z_{s}\cdots z_{w-1}\rangle\to
  |z_{s+1}\cdots z_{w-1}\rangle |k_4,y_{u+1}\cdots y_{u-1},z_{w+1}
  \cdots z_{s-1}\rangle.
  \]
\item Totally non-adjacent (TNA): Here the
  momentum from the other trace is $k_2$ and states transform like
  \[
  |y_u\cdots y_{u-1}\rangle |z_w\cdots z_{s}\cdots z_{w-1}\rangle\to
  |y_{u+1}\cdots y_{u-1},z_{w+1}
  \cdots z_{s-1}\rangle|k_4,z_{s+1}\cdots z_{w-1}\rangle. 
  \]
\item Totally adjacent (TA):
  \[
  |y_u\cdots y_{u-1}\rangle |z_w, z_{w+1}\cdots z_{w-1}\rangle\to
  |y_{u+1}\cdots y_{u-1}\rangle|k_4, z_{w+2}\cdots z_{w-1}\rangle. 
  \]
  Since $B(k_1,k_3)=-B(k_3,k_1)$, the permutation signs decide
  whether the term cancels or doubles.
\item Totally adjacent on the wrong side (TAWS) and
  momentum from the other trace is either $k_1$ or $k_3$:
  $(y_u=k_1,z_w=k_2,z_{w-1}=k_3)$
  and $(y_u=k_3,z_w=k_1,z_{w-1}=k_2)$. {\em Both} leading to  
  \[
  |y_u\cdots y_{u-1}\rangle |z_{w-1}, z_{w}\cdots z_{w-2}\rangle\to
  |z_{w+1}\cdots z_{w-2}\rangle |k_4, y_{u+1}\cdots y_{u-1}\rangle.
  \]
 \end{itemize}
On to the $1/N^3_c$ interactions.
It is not hard to generalize the $B2_{3Tr}$ interaction on three
traces \cite{UT6} to yield two instead of just one trace. Annihilating a parton
of the first trace by a non-planar, and from the other two traces by a planar
interaction, we join the latter two into one trace, and obtain a $1/N^3$
interaction ($1/N$ non-planar, $1/N$ parton loss, $1/N$ from
$g_{\rm YM}^2/4\pi$), with factors
\[
\left(-\frac{1}{N_c}\delta^{a_1}_{a_2}\delta^{i}_{j}\delta_{k_3}^{y^{(1)}_1}\right)
\left(\delta^{b_1}_{k}\delta^{b_2}_{l}\delta_{k_1}^{y_1^{(2)}}\right)
\left(\delta^{c_1}_{l}\delta^{c_2}_{i}\delta_{k_2}^{y^{(3)}_1}\right).
\]
Therefore the in and out-states look schematically like
\[
\widetilde{B3}_{23}:|y_u\cdots y_{u-1}\rangle|z_w\cdots z_s\cdots z_{w-1}\rangle
 \to
|y_{u+1}\cdots y_{u-1}\rangle|z_{w+1}\cdots z_{s-1}\rangle|k_4 z_{s+1}\cdots z_{w-1}\rangle.  
\]
{\small
\[
\overline{B3}_{32}:|y_u\cdots y_{u-1}\rangle|z_w\cdots z_{w-1}\rangle
|x_s\cdots x_{s-1}\rangle\to
|y_{u+1}\cdots y_{u-1}\rangle|z_{w+1}\cdots z_{w-1} k_4 x_{s+1}\cdots x_{s-1}\rangle.
\]
}
We can represent these interactions by diagrams following the same conventions
as in \cite{UT6}. Essentially, the momenta (red, blue, green)
of two or three annihilated fermions are changed (pink, cyan)
such that their sum remains the same, and we keep
track of the head (circle) and tail (square) of substrings of fermion
operators.
   

\unitlength0.7cm

\begin{picture}(14,2)
  \thicklines
  \put(1.0,1){\color{black}\oval(1.5,1.5)}
  \put(1.0,1){\color{red}\thicklines\oval(1.55,1.55)[bl]}
  \put(1.0,1){\color{red}\thicklines\oval(1.5,1.5)[bl]}
  \put(1.0,1){\color{red}\thicklines\oval(1.45,1.45)[bl]}
  \put(1.0,1){\color{red}\thicklines\oval(1.4,1.4)[bl]}
  \put(0.25,1){\color{red}\linethickness{0.09cm}\framebox(0.001,0.001)}
  \put(1,0.25){\color{red}\thicklines\circle*{0.29}}

  \put(3,1){\color{black}\oval(1.5,1.5)}
  \put(3.0,1){\color{blue}\thicklines\oval(1.55,1.55)[bl]}
  \put(3.0,1){\color{blue}\thicklines\oval(1.5,1.5)[bl]}
  \put(3.0,1){\color{blue}\thicklines\oval(1.45,1.45)[bl]}
  \put(3.0,1){\color{blue}\thicklines\oval(1.4,1.4)[bl]}
  \put(3,0.25){\color{blue}\thicklines\circle*{0.29}}
  \put(2.25,1){\color{blue}\linethickness{0.09cm}\framebox(0.001,0.001)}
  
  \put(4.25,1){\vector(1,0){1}}

  \put(6.5,1){\color{black}\oval(1.5,1.5)}
  \put(6.365,0.25){\color{red}\linethickness{0.09cm}\framebox(0.001,0.001)}
  \put(6.635,0.25){\color{red}\thicklines\circle*{0.29}}

  \put(9.5,1){\color{black}\oval(3.5,1.5)}
  \put(8.5,0.25){\color{pink}\linethickness{0.09cm}\line(1,0){1}}
  \put(9.5,0.25){\color{cyan}\linethickness{0.09cm}\line(1,0){1}}
  \put(8.5,0.25){\color{blue}\linethickness{0.09cm}\framebox(0.001,0.001)}
  \put(10.635,0.25){\color{blue}\thicklines\circle*{0.29}}

  \put(14.0,0.8){{\Large $\underline{A2}_{22}\simeq\underline{D2}_{22}$}} 

\end{picture}

\begin{picture}(14,2)
  \thicklines
  \put(1.0,1){\color{black}\oval(1.5,1.5)}
  \put(1.0,1){\color{red}\thicklines\oval(1.55,1.55)[bl]}
  \put(1.0,1){\color{red}\thicklines\oval(1.5,1.5)[bl]}
  \put(1.0,1){\color{red}\thicklines\oval(1.45,1.45)[bl]}
  \put(1.0,1){\color{red}\thicklines\oval(1.4,1.4)[bl]}
  \put(0.25,1){\color{red}\linethickness{0.09cm}\framebox(0.001,0.001)}
  \put(1,0.25){\color{red}\thicklines\circle*{0.29}}

  \put(3.75,1){\color{black}\oval(3,1.5)}
  \put(3.0,1){\color{blue}\thicklines\oval(1.55,1.55)[bl]}
  \put(3.0,1){\color{blue}\thicklines\oval(1.5,1.5)[bl]}
  \put(3.0,1){\color{blue}\thicklines\oval(1.45,1.45)[bl]}
  \put(3.0,1){\color{blue}\thicklines\oval(1.4,1.4)[bl]}
  \put(3,0.25){\color{blue}\thicklines\circle*{0.29}}
  \put(2.25,1){\color{blue}\linethickness{0.09cm}\framebox(0.001,0.001)}
  \put(4.5,1){\color{green}\thicklines\oval(1.55,1.55)[br]}
  \put(4.5,1){\color{green}\thicklines\oval(1.5,1.5)[br]}
  \put(4.5,1){\color{green}\thicklines\oval(1.45,1.45)[br]}
  \put(4.5,1){\color{green}\thicklines\oval(1.4,1.4)[br]}
  \put(5.25,1){\color{green}\thicklines\circle*{0.29}}
  \put(4.5,0.25){\color{green}\linethickness{0.09cm}\framebox(0.001,0.001)}

  \put(6.25,1){\vector(1,0){1}}

  \put(8.5,1){\color{black}\oval(1.5,1.5)}
  \put(8.365,0.25){\color{green}\linethickness{0.09cm}\framebox(0.001,0.001)}
  \put(8.635,0.25){\color{blue}\thicklines\circle*{0.29}}

  \put(11.5,1){\color{black}\oval(3.5,1.5)}
  \put(10.5,0.25){\color{pink}\linethickness{0.09cm}\line(1,0){1}}
  \put(10.5,0.25){\color{red}\linethickness{0.09cm}\framebox(0.001,0.001)}
  \put(11.5,0.25){\color{green}\thicklines\circle*{0.29}}
  \put(12.365,0.25){\color{blue}\linethickness{0.09cm}\framebox(0.001,0.001)}
  \put(12.635,0.25){\color{red}\thicklines\circle*{0.29}}

  \put(14.0,0.8){{\Large $\underline{B2}_{22}$} (PA1)} 

\end{picture}

\begin{picture}(14,2)
  \thicklines
  \put(1.0,1){\color{black}\oval(1.5,1.5)}
  \put(1.0,1){\color{red}\thicklines\oval(1.55,1.55)[bl]}
  \put(1.0,1){\color{red}\thicklines\oval(1.5,1.5)[bl]}
  \put(1.0,1){\color{red}\thicklines\oval(1.45,1.45)[bl]}
  \put(1.0,1){\color{red}\thicklines\oval(1.4,1.4)[bl]}
  \put(0.25,1){\color{red}\linethickness{0.09cm}\framebox(0.001,0.001)}
  \put(1,0.25){\color{red}\thicklines\circle*{0.29}}

  \put(3.75,1){\color{black}\oval(3,1.5)}
  \put(3.0,1){\color{blue}\thicklines\oval(1.55,1.55)[bl]}
  \put(3.0,1){\color{blue}\thicklines\oval(1.5,1.5)[bl]}
  \put(3.0,1){\color{blue}\thicklines\oval(1.45,1.45)[bl]}
  \put(3.0,1){\color{blue}\thicklines\oval(1.4,1.4)[bl]}
  \put(3,0.25){\color{blue}\thicklines\circle*{0.29}}
  \put(2.25,1){\color{blue}\linethickness{0.09cm}\framebox(0.001,0.001)}
  \put(4.5,1){\color{green}\thicklines\oval(1.55,1.55)[br]}
  \put(4.5,1){\color{green}\thicklines\oval(1.5,1.5)[br]}
  \put(4.5,1){\color{green}\thicklines\oval(1.45,1.45)[br]}
  \put(4.5,1){\color{green}\thicklines\oval(1.4,1.4)[br]}
  \put(5.25,1){\color{green}\thicklines\circle*{0.29}}
  \put(4.5,0.25){\color{green}\linethickness{0.09cm}\framebox(0.001,0.001)}

  \put(6.25,1){\vector(1,0){1}}

  \put(8.5,1){\color{black}\oval(1.5,1.5)}
  \put(8.365,0.25){\color{blue}\linethickness{0.09cm}\framebox(0.001,0.001)}
  \put(8.635,0.25){\color{green}\thicklines\circle*{0.29}}

  \put(11.5,1){\color{black}\oval(3.5,1.5)}
  \put(10.5,0.25){\color{pink}\linethickness{0.09cm}\line(1,0){1}}
  \put(10.5,0.25){\color{green}\linethickness{0.09cm}\framebox(0.001,0.001)}
  \put(11.5,0.25){\color{red}\thicklines\circle*{0.29}}
  \put(12.365,0.25){\color{red}\linethickness{0.09cm}\framebox(0.001,0.001)}
  \put(12.635,0.25){\color{blue}\thicklines\circle*{0.29}}

  \put(14.0,0.8){{\Large $\underline{B2}_{22}$} (PA3)} 

\end{picture}

\begin{picture}(14,2)
  \thicklines
  \put(1.0,1){\color{black}\oval(1.5,1.5)}
  \put(1.0,1){\color{red}\thicklines\oval(1.55,1.55)[bl]}
  \put(1.0,1){\color{red}\thicklines\oval(1.5,1.5)[bl]}
  \put(1.0,1){\color{red}\thicklines\oval(1.45,1.45)[bl]}
  \put(1.0,1){\color{red}\thicklines\oval(1.4,1.4)[bl]}
  \put(0.25,1){\color{red}\linethickness{0.09cm}\framebox(0.001,0.001)}
  \put(1,0.25){\color{red}\thicklines\circle*{0.29}}

  \put(3.75,1){\color{black}\oval(3,1.5)}
  \put(3.0,1){\color{blue}\thicklines\oval(1.55,1.55)[bl]}
  \put(3.0,1){\color{blue}\thicklines\oval(1.5,1.5)[bl]}
  \put(3.0,1){\color{blue}\thicklines\oval(1.45,1.45)[bl]}
  \put(3.0,1){\color{blue}\thicklines\oval(1.4,1.4)[bl]}
  \put(3,0.25){\color{blue}\thicklines\circle*{0.29}}
  \put(2.25,1){\color{blue}\linethickness{0.09cm}\framebox(0.001,0.001)}
  \put(4.5,1){\color{green}\thicklines\oval(1.55,1.55)[br]}
  \put(4.5,1){\color{green}\thicklines\oval(1.5,1.5)[br]}
  \put(4.5,1){\color{green}\thicklines\oval(1.45,1.45)[br]}
  \put(4.5,1){\color{green}\thicklines\oval(1.4,1.4)[br]}
  \put(5.25,1){\color{green}\thicklines\circle*{0.29}}
  \put(4.5,0.25){\color{green}\linethickness{0.09cm}\framebox(0.001,0.001)}

  \put(6.25,1){\vector(1,0){1}}

  \put(9,1){\color{black}\oval(2.5,1.5)}
  \put(8.335,0.25){\color{green}\linethickness{0.09cm}\framebox(0.001,0.001)}
  \put(8.665,0.25){\color{red}\thicklines\circle*{0.29}}
  \put(9.365,0.25){\color{red}\linethickness{0.09cm}\framebox(0.001,0.001)}
  \put(9.635,0.25){\color{blue}\thicklines\circle*{0.29}}

  \put(12.0,1){\color{black}\oval(2.5,1.5)}
  \put(11.5,0.25){\color{pink}\linethickness{0.09cm}\line(1,0){1}}
  \put(11.5,0.25){\color{blue}\linethickness{0.09cm}\framebox(0.001,0.001)}
  \put(12.5,0.25){\color{green}\thicklines\circle*{0.29}}

  \put(14.0,0.8){{\Large $\underline{B2}_{22}$} (TNA)} 

\end{picture}

\begin{picture}(14,2)
  \thicklines
  \put(1.0,1){\color{black}\oval(1.5,1.5)}
  \put(1.0,1){\color{red}\thicklines\oval(1.55,1.55)[bl]}
  \put(1.0,1){\color{red}\thicklines\oval(1.5,1.5)[bl]}
  \put(1.0,1){\color{red}\thicklines\oval(1.45,1.45)[bl]}
  \put(1.0,1){\color{red}\thicklines\oval(1.4,1.4)[bl]}
  \put(0.25,1){\color{red}\linethickness{0.09cm}\framebox(0.001,0.001)}
  \put(1,0.25){\color{red}\thicklines\circle*{0.29}}

  \put(3.75,1){\color{black}\oval(3,1.5)}
  \put(3.0,1){\color{blue}\thicklines\oval(1.55,1.55)[bl]}
  \put(3.0,1){\color{blue}\thicklines\oval(1.5,1.5)[bl]}
  \put(3.0,1){\color{blue}\thicklines\oval(1.45,1.45)[bl]}
  \put(3.0,1){\color{blue}\thicklines\oval(1.4,1.4)[bl]}
  \put(2.25,1){\color{blue}\linethickness{0.09cm}\framebox(0.001,0.001)}
  \put(3.0,0.25){\color{green}\linethickness{0.09cm}\line(1,0){1}}
  \put(4,0.25){\color{blue}\thicklines\circle*{0.29}}

  \put(6.25,1){\vector(1,0){1}}

  \put(8.5,1){\color{black}\oval(1.5,1.5)}
  \put(8.365,0.25){\color{red}\linethickness{0.09cm}\framebox(0.001,0.001)}
  \put(8.635,0.25){\color{red}\thicklines\circle*{0.29}}

  \put(11.0,1){\color{black}\oval(2.5,1.5)}
  \put(10.5,0.25){\color{pink}\linethickness{0.09cm}\line(1,0){1}}
  \put(10.5,0.25){\color{blue}\linethickness{0.09cm}\framebox(0.001,0.001)}
  \put(11.5,0.25){\color{blue}\thicklines\circle*{0.29}}

  \put(14.0,0.8){{\Large $\underline{B2}_{22}$} (TA)} 

\end{picture}

\begin{picture}(14,2)
  \thicklines
  \put(1.0,1){\color{black}\oval(1.5,1.5)}
  \put(1.0,1){\color{red}\thicklines\oval(1.55,1.55)[bl]}
  \put(1.0,1){\color{red}\thicklines\oval(1.5,1.5)[bl]}
  \put(1.0,1){\color{red}\thicklines\oval(1.45,1.45)[bl]}
  \put(1.0,1){\color{red}\thicklines\oval(1.4,1.4)[bl]}
  \put(0.25,1){\color{red}\linethickness{0.09cm}\framebox(0.001,0.001)}
  \put(1,0.25){\color{red}\thicklines\circle*{0.29}}

  \put(3.75,1){\color{black}\oval(3,1.5)}
  \put(3.0,1){\color{blue}\thicklines\oval(1.55,1.55)[bl]}
  \put(3.0,1){\color{blue}\thicklines\oval(1.5,1.5)[bl]}
  \put(3.0,1){\color{blue}\thicklines\oval(1.45,1.45)[bl]}
  \put(3.0,1){\color{blue}\thicklines\oval(1.4,1.4)[bl]}
  \put(2.25,1){\color{blue}\linethickness{0.09cm}\framebox(0.001,0.001)}
  \put(3.0,0.25){\color{green}\linethickness{0.09cm}\line(1,0){1}}
  \put(4,0.25){\color{blue}\thicklines\circle*{0.29}}

  \put(6.25,1){\vector(1,0){1}}

  \put(8.5,1){\color{black}\oval(1.5,1.5)}
  \put(8.365,0.25){\color{blue}\linethickness{0.09cm}\framebox(0.001,0.001)}
  \put(8.635,0.25){\color{blue}\thicklines\circle*{0.29}}

  \put(11.0,1){\color{black}\oval(2.5,1.5)}
  \put(10.5,0.25){\color{pink}\linethickness{0.09cm}\line(1,0){1}}
  \put(10.5,0.25){\color{red}\linethickness{0.09cm}\framebox(0.001,0.001)}
  \put(11.5,0.25){\color{red}\thicklines\circle*{0.29}}

  \put(14.0,0.8){{\Large $\underline{B2}_{22}$} (TAWS)} 

\end{picture}

\begin{picture}(18,2)
  \thicklines

  \put(1,1){\color{black}\oval(2.5,1.5)}
  \put(4.75,1){\color{black}\oval(4.5,1.5)}
  \put(3.25,0.25){\color{blue}\linethickness{0.09cm}\line(1,0){1}}
  \put(4.25,0.25){\color{blue}\thicklines\circle*{0.29}}
  \put(3.25,0.25){\color{blue}\linethickness{0.09cm}\framebox(0.001,0.001)}
  \put(1.5,0.25){\color{red}\thicklines\circle*{0.29}}
  \put(0.5,0.25){\color{red}\linethickness{0.09cm}\line(1,0){1}}
  \put(0.5,0.25){\color{red}\linethickness{0.09cm}\framebox(0.001,0.001)}
  \put(6.25,0.25){\color{green}\thicklines\circle*{0.29}}
  \put(5.25,0.25){\color{green}\linethickness{0.09cm}\line(1,0){1}}
  \put(5.25,0.25){\color{green}\linethickness{0.09cm}\framebox(0.001,0.001)}
  
  \put(7.75,1){\vector(1,0){1}}

  \put(10,1){\color{black}\oval(1.5,1.5)}
  \put(9.885,0.25){\color{red}\linethickness{0.09cm}\framebox(0.001,0.001)}
  \put(10.15,0.25){\color{red}\thicklines\circle*{0.29}}

  \put(12,1){\color{black}\oval(1.5,1.5)}
  \put(11.87,0.25){\color{green}\linethickness{0.09cm}\framebox(0.001,0.001)}
  \put(12.125,0.25){\color{blue}\thicklines\circle*{0.29}}

  \put(14.5,1){\color{black}\oval(2.5,1.5)}
  \put(14,0.25){\color{pink}\linethickness{0.09cm}\line(1,0){1}}
  \put(14,0.25){\color{blue}\linethickness{0.09cm}\framebox(0.001,0.001)}
  \put(15,0.25){\color{green}\thicklines\circle*{0.29}}

  \put(16.8,0.8){\Large $\widetilde{B3}_{23}$} 

\end{picture}

\begin{picture}(14,2)
  \thicklines
  \put(1.0,1){\color{black}\oval(1.5,1.5)}
  \put(1.0,1){\color{red}\oval(1.5,1.5)[br]}
  \put(1.0,1){\color{red}\oval(1.45,1.45)[br]}
  \put(1.0,1){\color{red}\oval(1.4,1.4)[br]}
  \put(1.0,1){\color{red}\oval(1.55,1.55)[br]}
  \put(1.75,1){\color{red}\thicklines\circle*{0.29}}
  \put(1,0.25){\color{red}\linethickness{0.09cm}\framebox(0.001,0.001)}
  
  \put(3.0,1){\color{black}\oval(1.5,1.5)}
  \put(3.0,1){\color{blue}\thicklines\oval(1.55,1.55)[bl]}
  \put(3.0,1){\color{blue}\thicklines\oval(1.5,1.5)[bl]}
  \put(3.0,1){\color{blue}\thicklines\oval(1.45,1.45)[bl]}
  \put(3.0,1){\color{blue}\thicklines\oval(1.4,1.4)[bl]}
  \put(2.25,1){\color{blue}\linethickness{0.09cm}\framebox(0.001,0.001)}
  \put(3,0.25){\color{blue}\thicklines\circle*{0.29}}

  \put(5.0,1){\color{black}\oval(1.5,1.5)}
  \put(5.0,1){\color{green}\thicklines\oval(1.55,1.55)[bl]}
  \put(5.0,1){\color{green}\thicklines\oval(1.5,1.5)[bl]}
  \put(5.0,1){\color{green}\thicklines\oval(1.45,1.45)[bl]}
  \put(5.0,1){\color{green}\thicklines\oval(1.4,1.4)[bl]}
  \put(4.25,1){\color{green}\linethickness{0.09cm}\framebox(0.001,0.001)}
  \put(5,0.25){\color{green}\thicklines\circle*{0.29}}
  
  \put(6.25,1){\vector(1,0){1}}

  \put(8.5,1){\color{black}\oval(1.5,1.5)}
  \put(8.375,0.25){\color{red}\linethickness{0.09cm}\framebox(0.001,0.001)}
  \put(8.675,0.25){\color{red}\thicklines\circle*{0.29}}

  \put(11.5,1){\color{black}\oval(3.5,1.5)}
  \put(10.5,0.25){\color{green}\linethickness{0.09cm}\framebox(0.001,0.001)}
  \put(10.75,0.25){\color{blue}\thicklines\circle*{0.29}}
  \put(11.5,0.25){\color{pink}\linethickness{0.09cm}\line(1,0){1}}
  \put(11.5,0.25){\color{blue}\linethickness{0.09cm}\framebox(0.001,0.001)}
  \put(12.5,0.25){\color{green}\thicklines\circle*{0.29}}

  \put(14.5,0.7){\Large $\overline{B3}_{32}$} 

\end{picture}


We should not be too surprised to find true multi-trace interactions.
For one, the multiparticle states of the theory should be expressible as
single-particle bound states, which mixes single-trace states,
i.e.~single-particle-state candidates, and multi-trace states.
Then there are the finite $N_c$ {\em null relations}
worked out in Ref.~\cite{KlebFiniteN},
which express multi-trace as single-trace states, e.g.~Eq.~(3.9) which implies
\beq\label{NullRelation}
\Tr\{B^{\dagger}(1)B^{\dagger}(1)B^{\dagger}(3)B^{\dagger}(3)\}|0\rangle=
\Tr\{B^{\dagger}(1)B^{\dagger}(3)\}\Tr\{B^{\dagger}(1)B^{\dagger}(3)\}|0\rangle,
\eeq
at $K=8$. In other words,
a one-trace state is identified with a two-trace state (at $N_c=2$).
Thirdly, the
ortho\-gonal eigenstates of the full Hamiltonian will have single- and
multi-trace components.

This is related to the Kutasov-Schwimmer universality \cite{KutasovSchwimmer},
which asserts that only adjoint and identity blocks can give rise to
single-particle states. Furthermore, the full Hamiltonian has to decompose
into Kac-Moody blocks (at least for $m=0$),
so we need interactions between the multi-trace states to make this happen.
If we restrict ourselves to the states in the identity and
adjoint blocks, which we can approximately identify with the single-trace block,
then certain interactions are absent. Therefore there is a chance that
this difference in available interactions sets the scale for the
$N_c$-invariance breaking of adjoint QCD$_2$. Unfortunately, we have not been
able to quantitatively verify this.
Indeed, most of the interactions, including the true multi-trace interactions,
are trivial
in the sense that they are only necessary to insure correct Kac-Moody
block decomposition. If we could somehow ``rotate'' them away by finding linear combinations of states that transform trivially under the Kac-Moody
Hamiltonian, we would be
left with the interactions of the true degrees of freedom, and would
have solved for the {\em bona fide} bound-state spectrum of the theory.
This is in a sense what the bosonization of fermion pairs into Kac-Moody
currents achieves \cite{Witten79}.
Unfortunately, even when transformation into non-orthogonal
states is implemented, there are still residual interactions\footnote{At least at finite harmonic resolution.} between multi-
and single particle states, although the multi-particle states with exact
multi-particle masses \cite{GHK} are projected out \cite{UT1}.

Looking at the null relation, Eq.~(\ref{NullRelation}), we might choose to view
the state as a one- or two-trace state, and when the Hamiltonian acts on
the state, the result has to be the same, irrespective of our choice.
Hence, there must be a symmetry of the Hamiltonian under this choice.
But the null relation is different for different $N_c$, so this must be a
very delicate symmetry.


\section{Deformations}

With all interactions in place, we can now run an enhanced DLCQ calculation,
i.e.~one which
allows us to choose the individual strengths of the interaction terms.
While these are no physical deformations --- in the sense that we
would add a term to the Lagrangian --- they give us a sense that the web of
interactions is intricate, and that some masses depend more than others
on certain interactions.

The motivation behind considering deformations is that the very weak $N_c$
dependence of the spectrum \cite{KlebFiniteN} may be made to disappear
altogether by adjusting the coupling strength
``vector'' $\vec{g}$, where the components represent different interaction
strengths.
Then there could be a path $\vec{g}(N_c)$ along which the mass spectrum would
be completely independent of $N_c$.

\begin{figure}[ht]
\centerline{
\psfig{file=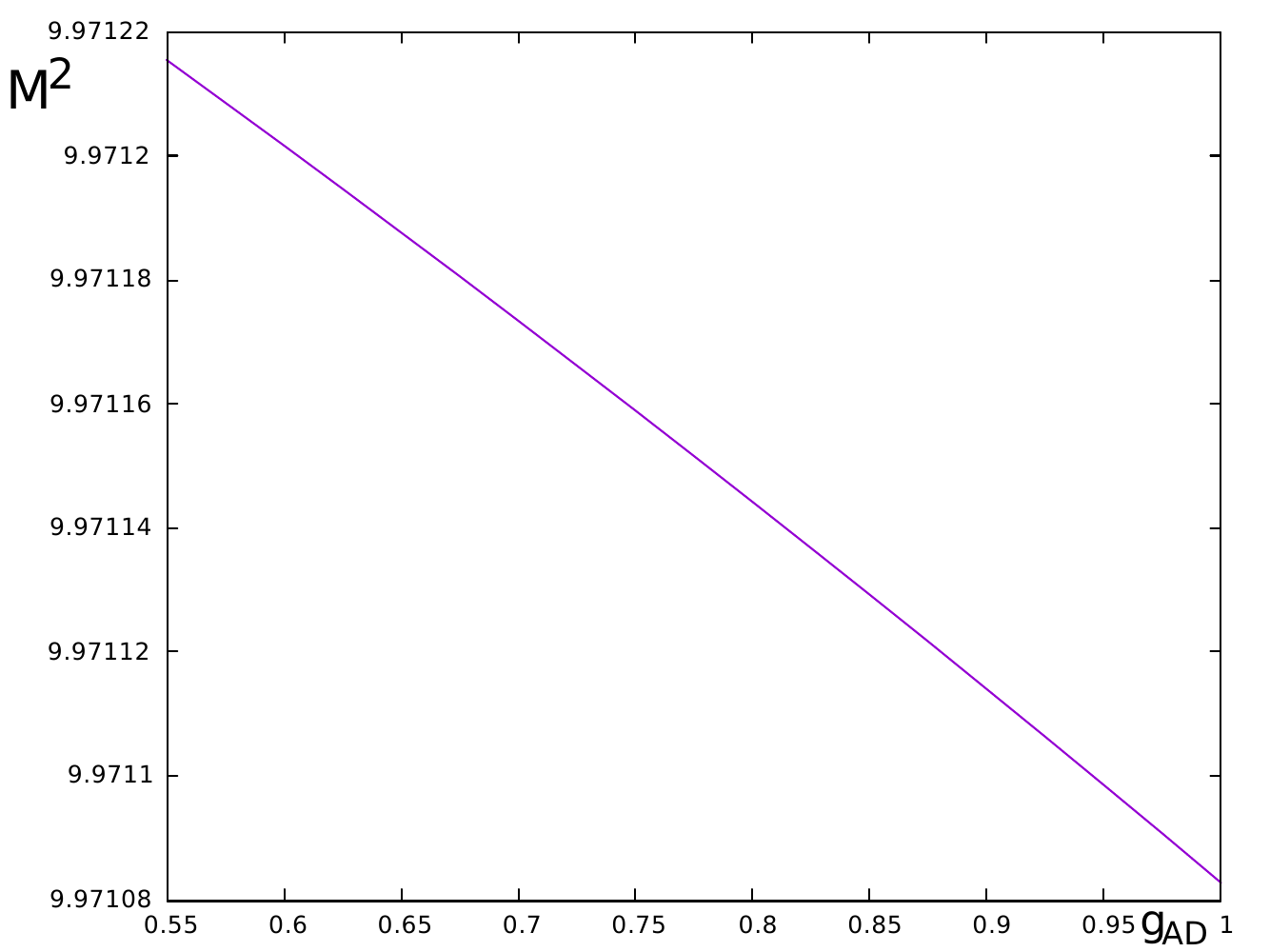,width=8cm}
\psfig{file=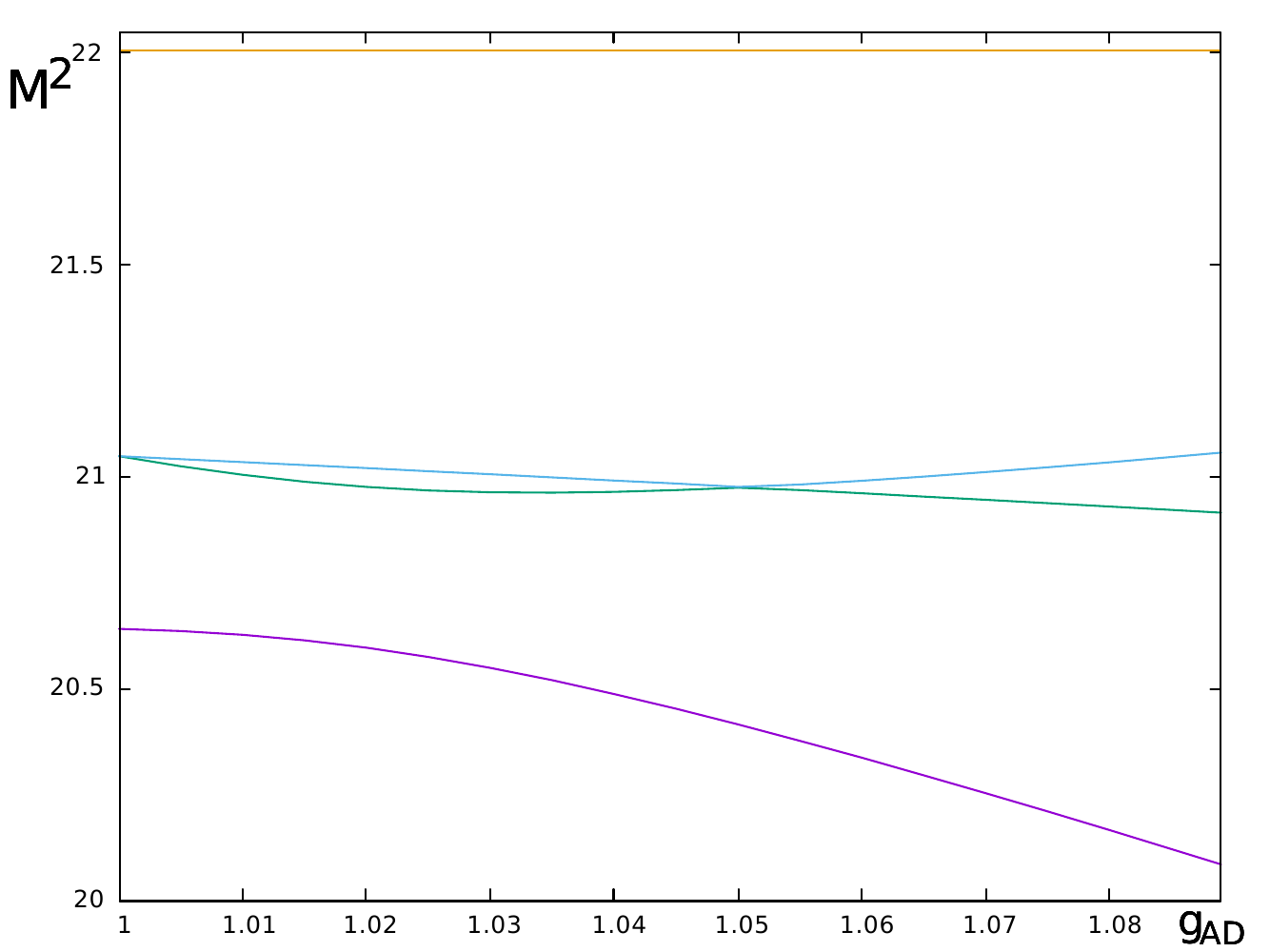,width=8cm}
}
\caption{Masses of low lying eigenstates of massless QCD$_{2A}$
  as a function of the $A1\& D1$ coupling  $g_{A1D1}$.
  As $g_{A1D1}$ increases, the $\underline{B2}$
  coupling  $g_{B_2}$ is weakened, so that $g_{A1D1}+g_{B2}=2$.}
\label{FigEVvsIAStrength}
\end{figure}

We computed the $N_c=2$ bound-state masses with
standard couplings, and then the $N_c=3$ masses with varying couplings, to see
if we can tune them to match masses at different $N_c$.
Leaving out some interactions wholesale leads to nonsensical results, so
we varied the interaction strengths very lightly. Typically
we dropped one by about ten percent and increased another by the same amount. 
The results are in Fig.~\ref{FigEVvsIAStrength}.
The left graph shows the lowest bound state mass of the massless theory
as a function of the $A1\& D1$ coupling\footnote{We identify trace-splitting and -joining couplings.}.
The ``undeformed'' masses are $M^2(N_c=2)=9.97116007$ and
$M^2(N_c=3)=9.97108279$ at $K=14$. We expect the behavior at higher resolutions
to be quantitatively but not qualitatively different.
We see that we can tune the lowest mass at $N_c=3$ by setting the coupling
strength to about 0.74. Of course, all other bound-state masses will also
change. This is dangerous, because we need to keep certain multi-particle
masses exactly degenerate, since their degeneracy is a consequence of
the Kac-Moody structure of the theory \cite{KlebFiniteN}.
Degeneracy is generally lost as couplings change,
but can be restored if other couplings are also allowed to
vary, as seen by the second
crossing of the two mass values on the right plot of
Fig.~\ref{FigEVvsIAStrength}, where both $A1\& D1$ and $\underline{B2}$
couplings are varied.
In general, though, the dependence of the bound-state masses on $N_c$ is too
intricate to be ``cured'' by adjusting a few coupling strengths.
As pointed out above, the theory and its Hilbert space  and null relations
look quite different at different $N_c$. While the $N_c$ dependence is
so weak that it suggests the existence of a symmetry,
the fact that $N_c$ is a discrete parameter is likely
making it impossible for the theory to be completely $N_c$ invariant.
We might speak of asymptotic $N_c$ invariance here when $N_c$ is large
but finite.

%
%

The original motivation that lead us to consider these (coupling) deformations
was the conjecture in Ref.~\cite{UT6} that some matrix element of
finite $N_c$ interactions completely vanish.
This gave hope to finding a
reason for the extremely weak dependence of the spectrum on $N_c$.
As will be described in Sec.~\ref{SecDivergence}, these hopes came
to naught. Nonetheless, it makes sense to ask whether some interactions are
more important than others. To this end, we studied the dependence of the
bound-state masses on individual couplings. We find a ``democratic'' situation,
where the changes of the masses as a function of the coupling
strengths, $M_j^2(g_i)$, are about equal for (at least)
the two most important ones
considered above, namely the parton-diagonal, trace off-diagonal interactions
$A1\& D1$ and the trace-diagonal, pair-annihilating
interaction $\underline{B2}$.


\section{The Divergence Structure and $N_c$ Dependence}
\label{SecDivergence}

Based on an observation in Ref.~\cite{Anton} that the contraction terms on
the Hamiltonian, Eq.~(\ref{ModeDecomp}), do not spawn finite
$N_c$ contributions,
we had argued in Ref.~\cite{UT6} that there must be a complete cancellation
between the singular part of the $A$-terms and the new finite $N_c$
singular $D$-terms. Specifically, we had conjectured that the
singular $1/N_c$ matrix elements vanish identically between
eigenstates $|\Psi^{(t)}_{\vec{r}}\rangle$ of the full theory.
Initially, we were encouraged to find
between states of $t$ traces and $r$ partons  
\beas
\langle t,r|\overline{A1}+\overline{D1}|t+1,r\rangle \approx
0,\qquad\qquad \langle t+1,r|\widetilde{A1}+\widetilde{D1}|t,r\rangle
\approx 0,
\eeas
where the approximation sign suggests that the
matrix elements of the sums $\overline{A1}+\overline{D1}$ and
$\widetilde{A1}+\widetilde{D1}$ are much smaller than the individual
matrix elements, but do not vanish altogether.
Alas, numerical results
using a hybrid method (discretized matrix elements in an eLCQ approach)
indicate that the conjecture is wrong. Namely, the matrix
elements of the operators between physical states are not zero in
general
\beas \langle
\Psi^{(t)}_{\vec{r}}|\overline{A1}+\overline{D1}|\Psi^{(t+1)}_{\vec{r}}\rangle
\neq 0,\qquad\qquad \langle
\Psi^{(t+1)}_{\vec{r}}|\widetilde{A1}+\widetilde{D1}|\Psi^{(t)}_{\vec{r}}
\rangle \neq 0.
\eeas
Indeed, technically we
only need the matrix elements to cancel at the location of the
singularities, not everywhere in the block, as was the case with other
operators, e.g.~$\langle
t,r|\underline{A2}+\underline{D2}|t,r\rangle=0$.  However, since in
the continuum limit we have smooth functions, we would expect the
matrix elements 
to partially cancel in the vicinity
of these points. Furthermore, the change from a single-momentum basis to
an asymptotic eigenstate basis requires summation or
rather integration over all states and hence entire subspaces of the
$r$ parton integral domain. On the other hand, since there is the
spectator delta function, we would expect the matrix elements to be
zero over most of the domain, in particular where the spectators are
different. Despite these arguments, we find that the matrix elements 
in a block do not necessarily vanish altogether.

\subsection{Why is the $N_c$-dependence so weak?}

Having ruled out the possibility of empty Hamiltonian blocks as an
explanation of weak $N_c$ dependence, let's focus on 
the peculiar behavior of the interaction $\underline{B2}$. In
particular the vanishing of its $2\to 4$ parton matrix element motivates us to
estimate the effect of finite $N_c$ terms in general on the masses
of the lowest states of the theory.
The procedure follows Ref.~\cite{KlebFiniteN},
where DLCQ numerics and independently
Eigenvalue Perturbation Theory was used to show that the corrections
to the eigenvalues are of order $a_1/N_c^2$ with very small coefficients
$a_1\in[0.004,0.046]$. In the notation\footnote{Actually, we are using the
  newer convention $g_{\rm YM}=2g$ and we added ellipses to $P^-$ due to
  order $1/N^3$ interactions between multi-trace states, see above.}
of \cite{KlebFiniteN}, the matrix for $P^-$ at fixed $K$, $P^-_K$,
as well as the eigenvalues $E_i$ are expanded in powers of $1/N_c$
\[
P^-_K=\frac{g_{YM}^2N_c}{2\pi}\left(P^-_{0,K}+\frac{1}{N_c}P^-_{1,K}
+\frac{1}{N^2}P^-_{2,K}+\ldots\right)
\]
and
\[
E_{i,K}=\frac{g_{YM}^2N_c}{2\pi}\left(E_{i,0,K}+\frac{1}{N_c}E_{i,1,K}
+\frac{1}{N^2}E_{i,2,K}+\ldots\right).
\]
The $E_{i,0,K}$ are the eigenvalues of $P^-_{0,K}$ at leading
order. However, the matrix $P^-_{0,K}$ is not
Hermitian, so one obtains
  left and right eigenvectors, $\langle l_{i,K}|$ and  $| r_{i,K}\rangle$,
  which are chosen to be orthogonal,
  $\langle l_{i,K}| r_{j,K}\rangle=\delta_{ij}$.
  The first order correction due to $P_{1,K}$ turns out to
  vanish.

The order $1/N^2_c$ contributions come from the first order corrections of
$P^-_{2,K}$ and second order corrections due to $P^-_{1,K}$
\[
E_{i,2,K}=\langle l_{i,K}|P^-_{2,K}|r_{i,K}\rangle
+\sum_{j\neq i}\frac{\langle l_{j,K}|P^-_{1,K}|r_{i,K}\rangle
  \langle l_{i,K}|P^-_{1,K}|r_{j,K}\rangle}{E_{j,0,K}-E_{i,0,K}}
\]
The coefficient $a_1$ is then defined as
\[
a_1:=\lim_{K\to\infty} (KE_{i,2,K}),
\]
and was computed in \cite{KlebFiniteN} numerically at increasing $K$.
Excellent agreement
with an independent $1/N$ extrapolation was found for the lowest four
single-particle bound state masses of the theory.

Given our classification of all possible interactions between single- and
multi-trace states above, we can go a bit further.
First, we can set formally $P^-_{1,K}\sim A1+D1$ and
$P^-_{2,K}\sim \underline{B2}$. Next, we repeat our finding from \cite{UT6}
that the only trace-diagonal finite $N_c$ correction comes from
$\underline{B2}$, but that $\langle r'=2|\underline{B2}|r=4\rangle =0$ and
also $\langle r'=4|P^-|r=2\rangle =0$. Since $A1$ and $D1$ are parton-number
preserving interactions, we also have
$\langle r'=2|\overline{A1}+\overline{D1}|r_1=2\rangle |r_2=2\rangle =0$.
In other words, there are (almost) no finite $N_c$ corrections to two-parton
states. On the other hand, parton-number purity \cite{DalleyKlebanov} implies
that the lowest bosonic state is an almost pure two-parton
state\footnote{The lowest fermionic eigenstate is an almost pure three-parton
  state.} (97\%). This means that corrections to the lowest mass eigenvalue
are due to its four-parton contributions, and therefore suppressed by
a factor $3\%=0.03$.

Note that the analysis in
Ref.~\cite{KlebFiniteN} is more general:
{\em all} $1/N$ or $1/N^2$ interactions are taken into account.
Our aim here is to find the reason for the very weak $N_c$ dependence:
why should the coefficients be so small? The simplest
explanation, namely that it is the suppression due to parton-purity, is too
weak: it does not explain why the second-lightest boson at $M^2\approx 25.6$
also exhibits a small $a_1$ even though it is an almost pure four-parton
state with essentially zero two-parton contributions.
Hence, there must be another reason for the smallness of the
$N_c$-dependence --- maybe the symmetry alluded to above.
We have not been able to find such a symmetry, but will rule out certain
candidates in the next subsection.

\subsection{Cautionary Remarks on Symmetries}

It seems counter-intuitive to look for advice on two-dimensional QCD
in four-dimension\-al QCD, but here we go.
In an interesting paper \cite{Yuri2009}, the authors show
that a suppression of
finite $N_c$ effects happens also in full QCD. While the paper
assumes very specific circumstances, it nonetheless finds effects
{\em fixed by group theory constraints} that are responsible for a 100-fold
reduction of $N_c$-sensitivity in 
the small-$x$ evolution in the Color Glass Condensate picture beyond the
Balitsky-Kovchegov truncation, which is a large $N_c$ approximation
of sorts.

This suggests that it is possible to find --- for
the simpler theory at hand --- a 
group theoretical argument as to why the $N_c$-sensitivity of adjoint
QCD$_2$ is so small. A straightforward line of reasoning
goes as follows. Start with the dilemma in Ref.~\cite{Kutasov94}:
To discuss pair production, the asymptotic wavefunctions
$\phi_{2k}$ constructed therein
are insufficient, because their $\mathbb{Z}_2$ charge under
the symmetry ${\cal T}$, Eq.~(\ref{Tsymmetry})
changes with parton number, namely
$T\phi_{2k}=(-)^{k+1}\psi_{2k}$. Therefore, the $T$-symmetric Hamiltonian
cannot have matrix elements between these states if their parton number
differs by two, $\langle\phi_{2k}|P^-|\phi_{2k\pm 2}\rangle=0$.
New (asymptotic) eigenfunctions $\tilde{\phi}_k$ with $T=(-)^k$
are necessary. These were constructed in \cite{UT3and4,UT5}, and their $TIS$
quantum numbers under the symmetries of the theory appear in Table 1
of Ref.~\cite{UT5}. From this table it can be gleaned that the quantum numbers
of states with $r$ partons are the same as those of states with parton number 
$r\pm\Delta r$, where $\Delta r=4s$, $s\in\mathbb{N}$.
However, the Hamiltonian preserves $T_s$ not $T$, as can be
gleaned from the fact that the Hamilton matrix we are diagonalizing
has matrix elements between {\em states} of the same $T_s$. This
means that the above argument is flawed:
$\langle\phi_{2k}|P^-|\phi_{2k\pm 2}\rangle\neq 0$,
as the wavefunctions Eq.~(4.13) and (4.15) of \cite{Kutasov94},
i.e.~$\phi_4(4,2)=\varphi_{4-}(4,0,2)$ and  
$\phi_6(6,4,2)=\varphi_{6+}(6,0,4,0,2)$, are of opposite $T$ but the
associated states are both $T_s$ odd.
Thus, the question of pair production {\em could} have been addressed.

Apropos, $T_s$ is the only quantum number that can be used for a 
superselection-type argument.
The other three ``symmetries'' ${\cal C},{\cal I}, {\cal S}$
effectively act as constraints.
For instance, due to the anti-commutation relation of the quarks, we
need bosonic(fermionic) wavefunctions (not) to change sign\footnote{This is
  just a {\em consistency constraint}. For instance, since
\beas
|\Psi_4\rangle&=&\int\frac{d[x_4]}{N_c^2}\psi(x_1,x_2,x_3,x_4)\Tr\{b^\dagger(x_1)b^\dagger(x_2)b^\dagger(x_3)b^\dagger(x_4)\}\\
&=&-\int \frac{d[x_4]}{N_c^2} \psi(x_1,x_2,x_3,x_4)\Tr\{b^\dagger(x_2)b^\dagger(x_3)b^\dagger(x_4)b^\dagger(x_1)\},
\eeas
we have to identify $\psi(x_2,x_3,x_4,x_2)=-\psi(x_1,x_2,x_3,x_4)$,
i.e.~we need a wavefunction antisymmetric under cyclic permutation of
its arguments. Note that we cannot rename the momenta, since the integration
measure, Eq.~(\ref{IntMeasure}), fixes the uHS cell. } 
under ${\cal C}$, $\phi_r(x_1,\ldots, x_r)=(-1)^{r-1}\phi_r(x_2,\ldots, x_r,x_1)$.
The reality of the quarks forces the eigen{\em functions} to be
even or odd under inversion ${\cal I}$,
and the ``boundary conditions''
(better: the vanishing or extremalization of the wavefunction at the boundary)
requires {\em definite} ${\cal S}$ parity. 
Really all the eLCQ group theory \cite{UT5}
does is to assign appropriate $I$ and $S$
quantum numbers to the parton sectors, since $C=(-1)^{r-1}$ is fixed. 

Hence, there is no ``superselective'' way to decide whether the
matrix element of
$N_c$-symmetry breaking operators like $\underline{B2}$ vanish between
asymptotic states, and a symmetry that would explain the smallness of the
$N_c$ dependence of the spectrum must be of a different type.


\section{Conclusions}

In this short note we remarked on several aspects of the parameter dependence
of QCD$_{2A}$.
For the large $N_c$ theory we showed that increasing constituent mass will
drive the eigenstates towards purity in parton number due to the suppression
of pair creation.
For the finite $N_c$ theory the set of interactions was completed by adding
to the list in Ref.~\cite{UT6} five operators with matrix elements between
two multi-particle states. We pointed out that the existence of null relations
\cite{KlebFiniteN} at finite $N_c$ manifestly connects single- and multi-trace
states by identifying them. This introduces an ambiguity into the action
of the Hamiltonian on such states, which could hint at a symmetry of the theory.
The null relations are different at different $N_c$, yet the spectrum of the
theory is almost insensitive to $N_c$, which seems curious.

We used the completed list of interactions to deform the theory by
varying couplings of individual operators. The ensuing mass spectrum
showed us that these operators are about equally important to ensure that
certain multi-particle masses are degenerate, as mandated by the Kac-Moody
structure of the (massless) Hamiltonian. The varying of couplings allowed
us to dial in complete $N_c$ invariance of some bound state masses. However,
we found that masses reacted to changed coupling {\em individually}, which
virtually rules out the possibility of being able to adjust a few
couplings to render even a significant fraction of the states' masses
$N_c$-invariant.

Going back to large $N_c$, it would be interesting to treat fundamentals
in the eLCQ framework, although it is not immediately clear how asymptotic
functions would look like, or if they are linearly independent of the eLCQ
basis states. Just as fundamentals ``fix'' the ends of the closed strings
representing glueballs, we should open up a new class of functions to
accommodate these new solutions.

Due to the renewed efforts of the last couple of years
to understand the spectrum of two-dimensional
adjoint QCD with various and varied
approaches like DLCQ \cite{KlebFiniteN}, the lattice \cite{KlebLattice,
 Bergner2024},
analytic work \cite{Seifnashri,Jacobson}, and eLCQ \cite{UT3and4,UT5},
to name a few, we have a good idea what the spectrum looks like.
But if we are honest, one of the most important questions, as to what the true single
particle spectrum of the theory is, and whether there are
indeed --- as conjectured --- (infinitely) many Regge trajectories of bound
states, has not been answered in a satisfactory way.
Even with enhanced computing power and
sophisticated group theoretical methods at hand \cite{KlebFiniteN}, the DLCQ
spectrum is too riddled with multi-particle states to glean but a few of
the low single-particle states. Most of the latter are well hidden in multiple
multi-particle state multiplets. The asymptotic basis-function approach of eLCQ
has far fewer states, but cannot be run to sufficiently high parton numbers
to ensure convergence of states --- at least not in the massless limit. 
It would therefore be interesting to check the numerical
finite $N_c$ calculations \cite{KlebFiniteN} in the
bosonized or current picture. In fact, even large $N_c$ DLCQ calculations
of the bosonized theory could provide more clarity --- at the cost of dealing with
non-orthogonal bases and diagonalization of large asymmetric matrices.


\section*{Acknowledgments}

I'd like to thank the Ohio State University's Physics 
Department for hospitality while this work was being completed.
I thank Igor R.~Klebanov for suggesting work on the large constituent mass
regime and to study deformations of the theory. 





\begin{thebibliography}{99}
\bibitem{DalleyKlebanov}
  S.~Dalley and I.~R.~Klebanov,
  Phys.\ Rev.\ D {\bf 47} (1993) 2517.
\bibitem{BDK}
  G.~Bhanot, K.~Demeterfi and I.~R.~Klebanov,
  Phys.\ Rev.\ D {\bf 48} (1993) 4980.
\bibitem{Kutasov94}
         {D.~Kutasov},
         {\em Nucl.~Phys.~}{\bf B414}
         (1994)
         33.
\bibitem{GHK}
  D.~J.~Gross, A.~Hashimoto and I.~R.~Klebanov,
  Phys.\ Rev.\  D {\bf 57} (1998) 6420.
\bibitem{Gomis} D.~Delmastro, J.~Gomis and M.~Yu,
JHEP \textbf{02} (2023), 157; D.~Delmastro and J.~Gomis,
JHEP \textbf{09} (2023), 158.
\bibitem{Damia2024} J.~A.~Damia, G.~Galati, L.~Tizzano,
JHEP \textbf{12} (2025), 230.
\bibitem{Boorstein}
    {D.~Kutasov, J. Boorstein}, {\em Nucl.~Phys.~}{\bf B421} (1994) 263--277.
\bibitem{Schoutens} Gopakumar, A.~Hashimoto, I.~R.~Klebanov, S.~Sachdev,
      K.~Schoutens, Phys.~Rev.~D {\bf 86} (2012) 066033.
\bibitem{GrossKlebMatSmilga}
D.~J.~Gross, I.~R.~Klebanov, A.~V.~Matytsin and A.~V.~Smilga,
Nucl. Phys. B \textbf{461} (1996), 109-130.
\bibitem{Anton}
  F.~Antonuccio and S.~Pinsky,
  Phys.\ Lett.\  B {\bf 439} (1998) 142.
 \bibitem{Jacobson} A.~Cherman, T.~Jacobson, Y.~Tanizaki and M.~\"Unsal,
SciPost Phys. \textbf{8} (2020) no.5, 072.
\bibitem{KlebFundamental21}
 R.~Dempsey, I.~R.~Klebanov and S.~S.~Pufu,
JHEP \textbf{10} (2021), 096.
\bibitem{KlebSmallCircle}
 R.~Dempsey, I.~R.~Klebanov,S.~S.~Pufu, B.~T.~S\o{}gaard,
JHEP \textbf{11} (2024), 128.
\bibitem{Seifnashri}
Z.~Komargodski, K.~Ohmori, K.~Roumpedakis and S.~Seifnashri,
JHEP \textbf{03} (2021), 103.
\bibitem{Smilga21} A.~V.~Smilga,
SciPost Phys. \textbf{10} (2021) no.6, 152.
\bibitem{KutasovSchwimmer}
  D.~Kutasov and A.~Schwimmer,
  Nucl.\ Phys.\  B {\bf 442} (1995) 447.
\bibitem{UT1}
  U.~Trittmann,
  Nucl.\ Phys.\  B {\bf 587}, 311 (2000),
  Phys.\ Rev.\  D {\bf 66}, 025001 (2002).
      \bibitem{Katz}
  E.~Katz, G.~M.~Tavares and Y.~Xu,
  JHEP {\bf 1405} (2014) 143.
\bibitem{Witten79}
  E.~Witten,
  {\em Nuovo Cim. A} {\bf 51} (1979) 325. 
\bibitem{KlebSupercurrents}
  I.R.~Klebanov, S.S.~Pufu, B.~S\o{}gaard, E.~Witten,
JHEP \textbf{04} (2026), 212.
\bibitem{HamLattice1and2}
    R.~Dempsey, I.R.~Klebanov, S.S.~Pufu, B.~S\o{}gaard,   JHEP 08 (2024) 009; 
    R.~Dempsey, S.S.~Pufu, B.~S\o{}gaard, I.R.~Klebanov,  JHEP 06 (2025) 260. 
\bibitem{DempseyGlueck} R.~Dempsey, A.-M.E.~Gl\"uck, S.S.~Pufu, B.~S\o{}gaard,
      arXiv:2601.16262 [hep-th].
\bibitem{UT3and4}
  U.~Trittmann, 
  Phys.\ Rev.\  D {\bf 92} (2015) 8, 085021;
  Phys.\ Rev.\  D {\bf 98},(2018) 4, 045011.
\bibitem{UT5}
  U.~Trittmann, 
  Phys.\ Rev.\  D {\bf 109} 1, 016004 (2024).
\bibitem{UT6}
  U.~Trittmann, 
  Phys.\ Rev.\  D {\bf 110} 6, 065023 (2024).
  \bibitem{KlebFiniteN}
  R.~Dempsey, I.~R.~Klebanov, L.~L.~Lin and S.~S.~Pufu,
JHEP \textbf{04} (2023), 107.
\bibitem{Asrat} M.~Asrat, Phys.~Rev.~D {\bf 107}, 106022 (2023).  
\bibitem{KlebLattice}
R.~Dempsey, I.~R.~Klebanov, S.~S.~Pufu and B.~T.~S\o{}gaard,
JHEP 08 (2024) 009.
\bibitem{Bergner2024}
G.~Bergner, S.~Piemonte and M.~\"Unsal,
JHEP 07 (2024) 048.
\bibitem{Komatsu} F.~Ambrosino and S.~Komatsu,
  JHEP 02 (2025) 126.
\bibitem{Yung}
  L.~Yung, ``Asymptotic spectrum of 2D adjoint QCD in the large $N$ limit'',
  master thesis (2022), Weizmann Institute, Rehovot, Israel (unpublished).
\bibitem{Aharony}
  O.~Aharony, private communication.
\bibitem{KlebPrivate}
  I.R.~Klebanov, private communication.
\bibitem{Yuri2009} Yu. V. Kovchegov, J.~Kuokkanen, K.~Rummukainen, H.~Weigert,
Nucl.~Phys.~{\bf A813} (2009) 47--82.
\end{thebibliography}
\end{document}